\documentclass[fleqn,usenatbib, useAMS]{mnras}

\usepackage{newtxtext,newtxmath}

\usepackage[T1]{fontenc}
\usepackage{ae,aecompl}
\usepackage[utf8]{inputenc}
\usepackage{threeparttable}
\usepackage{booktabs, caption, makecell}
\usepackage{pdflscape}
\usepackage{geometry}

\usepackage{float}
\usepackage{pifont}
\newcommand{\xmark}{\ding{53}}%
\newcommand{\Y}{\checkmark}%
\newcommand{\N}{\xmark}%

\usepackage{graphicx}	
\usepackage{amsmath}	

\newcommand{\Ion}[2]{#1{\,\sc#2}}

\usepackage{rotating}
\usepackage{float}
\usepackage{tabularx}
\usepackage{natbib}
\usepackage{hyperref}
\usepackage{academicons}
\usepackage{svg}
\newcommand{\orcid}[1]{\href{https://orcid.org/#1}{\includegraphics[width=10pt]{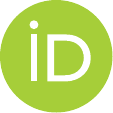}}}
\newcommand{\Teff}{\mbox{$T_{\mathrm{eff}}$}}
\newcommand{\logg}{\mbox{$\log g$}}
\newcommand{\Msun}{\mbox{$\mathrm{M_\odot}$}}
\newcommand{\totalnumber}{311}
\newcommand{\SIAM}{SIAM Journal on Numerical Analysis}

\title[\textit{HST} COS UV spectroscopic survey of DA white dwarfs]{An \textit{HST} COS ultra-violet spectroscopic survey of \totalnumber\ DA white dwarfs.\\
I. Fundamental parameters and comparative studies}
\author[Sahu et al.]
{Snehalata Sahu\orcid{0000-0002-0801-8745}$^{1}$\thanks{E-mail: snehalatash30@gmail.com}, 
Boris T. G\"ansicke\orcid{0000-0002-2761-3005}$^{1}$, 
Pier-Emmanuel Tremblay\orcid{0000-0001-9873-0121}$^{1}$, 
Detlev Koester\orcid{0000-0002-6164-6978}$^{2}$, 
\newauthor{J.J. Hermes\orcid{0000-0001-5941-2286}$^{3}$, 
David J. Wilson\orcid{0000-0001-9667-9449}$^4$, 
Odette Toloza\orcid{0000-0002-2398-719X}$^{5}$, 
Matthew J. Hoskin\orcid{0000-0003-3057-1886}$^{1}$, 
Jay Farihi\orcid{0000-0003-1748-602X}$^{6}$,}
\newauthor{Christopher J. Manser\orcid{0000-0003-1543-5405}$^{7}$, 
Seth Redfield\orcid{0000-0003-3786-3486}$^8$}\\\\
$^{1}$ Department of Physics, University of Warwick, Coventry, CV4 7AL, UK\\
$^{2}$ Institut für Theoretische Physik und Astrophysik, University of Kiel, 24098 Kiel, Germany\\
$^{3}$Department of Astronomy \& Institute for Astrophysical Research, Boston University, 725 Commonwealth Ave., Boston, MA 02215, USA\\
$^4$Laboratory for Atmospheric and Space Physics, University of Colorado, Boulder, CO 80303\\
$^{5}$Departamento de Física, Universidad Técnica Federico Santa María, Avenida España 1680, Valparaíso, Chile\\
$^{6}$Department of Physics \& Astronomy, University College London, Gower Street, London WC1E 6BT, UK\\
$^{7}$Astrophysics Group, Department of Physics, Imperial College London, Prince Consort Rd, London, SW7 2AZ, UK\\
$^{8}$Astronomy Department and Van Vleck Observatory, Wesleyan University, Middletown, CT 06459-0123, USA
}
\date{Accepted 2023 August 30. Received 2023 August 29; in original form 2023 July 24}

\begin{document}
\label{firstpage}
\pagerange{\pageref{firstpage}--\pageref{lastpage}}
\maketitle

\begin{abstract}
White dwarf studies carry significant implications across multiple fields of astrophysics, including exoplanets, supernova explosions, and cosmological investigations. Thus, accurate determinations of their fundamental parameters (\Teff\ and \logg) are of utmost importance. While optical surveys have provided measurements for many white dwarfs, there is a lack of studies utilising ultraviolet (UV) data, particularly focusing on the warmer ones that predominantly emit in the UV range. Here, we present the medium-resolution far-UV spectroscopic survey of \totalnumber\ DA white dwarfs obtained with Cosmic Origins Spectrograph (COS) onbaord \textit{Hubble Space Telescope} confirming 49 photometric \textit{Gaia} candidates. We used 3D extinction maps, parallaxes, and hydrogen atmosphere models to fit the spectra of the stars that lie in the range $12\,000 < \Teff < 33\,000$\,K, and $7 \leq\logg < 9.2$. To assess the impact of input physics, we employed two mass-radius relations in the fitting and compared the results with previous studies. The comparisons suggest the COS \Teff\ are systematically lower by 3\,per cent on average than Balmer line fits while they differ by only 1.5\,per cent from optical photometric studies. The mass distributions indicate that the COS masses are smaller by $\approx0.05$\,\Msun\ and 0.02\,\Msun\ than Balmer lines and photometric masses, respectively. Performing several tests, we find that the discrepancies are either arising due to issues with the COS calibration, broadening theories for hydrogen lines, or interstellar reddening which needs further examination. Based on comparative analysis, we identify 30 binary candidates drawing attention for follow-up studies to confirm their nature.
\end{abstract}


\section{Introduction}
The fundamental parameters such as effective temperatures and surface gravities serve as foundation stones for scientific studies related to the field of white dwarfs. A measure of \Teff\ and \logg\ is essential for determining their masses, radii, ages, and luminosities. Thus, characterising a sufficiently large sample of white dwarfs is key for studying their mass distribution, which holds insights into the formation of single and binary stars \citep{Finley1997, Bergeron1992, Kepler2007, Tremblay2016}. White dwarfs are also crucial in constraining the initial-to-final mass relation  \citep{Kurtis2004, Raddi2016, Cummings2018} that is vital in the context of mass-loss throughout the stellar evolution process as well as the star formation history in the solar neighbourhood \citep{Elena2023}. These studies have far-reaching implications, ranging from the exploration of exo-planetary systems \citep{boris2019,bonsor2023} to understanding supernova explosions \citep{supernova, Greiner:2023viw} to contributing to cosmological investigations \citep{doi:10.1126/science.abd1714}. 

The majority of white dwarfs known in our Galaxy (up to 80\,per cent) are of DA spectral type whose spectra at optical wavelengths are dominated by hydrogen (H) Balmer absorption lines. Their atmospheres have been modeled in great detail, resulting in the accurate derivation of their parameters specifically in the optical bands. Extensive spectroscopic surveys have contributed to this understanding by providing parameters for large samples that involves fitting the Balmer lines with synthetic spectra based on atmospheric models \citep{Bergeron1992,Finley1997, Marsh1997,Koester2009,Gianninas2011,Tremblay2011,Kepler2019, genest2019}. Further, there exist several photometric studies conducted using various telescopes and surveys such as \textit{Gaia}, Pan-STARRS, and SDSS \citep{Tremblay2019_param, Bergeron2019, Kilic2020, nicola2021, Esteban2022} that have obtained the parameters by comparing the synthetic photometry with the observed magnitudes in the respective band-passes. These studies primarily cover the optical wavelength regions spanning from 3500 to 9300\,\AA. However, the derivation of fundamental parameters from other spectral regions has been less explored, for instance, \cite{Lajoie2007, Wall2023} using ultraviolet (UV) observations, \cite{nicola2020} using \textit{Hubble Space Telescope} (\textit{HST}) STIS + WFC3 and infrared observations. These investigations are crucial as they enable a comparison of parameters derived from different observational techniques. Such comparisons can aid in discerning the systematic data effects, uncovering the limitations in model atmospheres, and identifying intriguing objects, such as binary systems. By expanding the parameter derivation beyond the optical range, these studies contribute to a more comprehensive understanding of white dwarf properties and their diverse observational characteristics.

In this regard, UV observations are important, as the Lyman\,$\alpha$ (1216\,\AA) absorption line of hydrogen is the dominant feature in the UV spectra.
However, because white dwarfs are small and correspondingly intrinsically faint, only a limited number have been adequately observed in the UV. Consequently, only a small number of published studies have used UV data for the determination of the parameters.
For example, studies conducted during the 1980$-$2000s have used \textit{International Ultraviolet Explorer (IUE)} data covering the Ly$\alpha$ region to derive \Teff\ for a relatively small sample of DAs focusing on those hotter than 20\,000\,K \citep{Holberg1986} or pulsating white dwarfs spanning the temperature range 11\,000-13\,000\,K \citep{Kepler1993, Bergeron1995}. Using \textit{Far Ultraviolet Spectroscopic Explorer (FUSE)} observations of 16 DA white dwarfs, \cite{Barstow2003} found that the \Teff\ values obtained from Lyman lines are in reasonable agreement with the optical parameters derived from Balmer line fitting, showing deviations only for very hot stars ($>$50\,000\,K), also noted in \cite{Good2004}. Later, \cite{Lajoie2007} arrived at a similar conclusion based on their statistical comparison of UV and optical temperatures of a much larger sample of 140 objects using \textit{IUE} data. There are some detailed UV analyses available for individual stars using data from the \textit{Extreme UV Explorer} \citep{Dupuis2000}, and the \textit{HST} \citep{Koester2014, Wilson2019}, where UV variability and metal pollution have been detected. Some of these individual studies reported significant discrepancies between the parameters derived from UV and optical observations. However, since these studies lacked access to parallax measurements, \logg\ values are solely based on optical data. Consequently, this approach does not offer an independent estimation of all the UV parameters. 

Comparisons of multi-wavelength observations, especially UV with optical studies are crucial in revealing the existence of unresolved double degenerate binaries that are the possible progenitors of Type Ia supernovae \citep{Lajoie2007, 2015MNRAS.450.3966B, Wall2023}. This is supported by composite spectra simulations of white dwarf model atmospheres \citep{Lajoie2007, Tremblay2011}. Further, UV observations are useful in the study of white dwarf-main sequence (MS) binaries. In these binary systems, the optical spectra (Balmer lines) can be contaminated by the MS companion making it difficult to precisely measure the white dwarf parameters, which is otherwise simpler in UV where the flux is mainly dominated by the hotter component. UV spectroscopic studies are also sensitive in detecting the heavy metal lines that serve as direct signatures of planetary debris being accreted from discs around the white dwarfs \citep{boris2012,farihi2013}. Thus, precise determinations of \Teff\ and \logg\ are essential to obtain accurate metal abundances and study their correlation with the fundamental parameters \citep[\Teff, mass, and cooling age;][]{Koester2014}.

There are only a few studies in the UV \citep{Lajoie2007,Wall2023} that have carried out a systematic analysis to understand the effect of different methods, models, or observations in the white dwarf parameters. Here, we present a far-UV spectroscopic survey of \totalnumber\ DA white dwarfs observed with the \textit{HST} Cosmic Origins Spectrograph (COS) from 2010 to 2023. 
Owing to the large number of DAs observed with \textit{HST} data, we planned to make a series of publications focusing on various science cases. In this paper (paper I), we conduct a comprehensive comparison of the fundamental parameters obtained using \textit{HST} UV observations with the previous photometric and spectroscopic studies, with the aim to assess the systematics and identify the potential sources of discrepancies. The COS spectra cover the UV spectral region, including Ly$\alpha$, thus, providing an excellent opportunity to precisely measure \Teff\ and \logg\ and test the accuracy of optically derived values. The targets studied in our survey lie in the intermediate temperature range  (12\,000 to 33\,000\,K) unlike previous UV studies that were mostly focused on hot white dwarfs ($\Teff>50\,000$\,K).

The outline of the paper is as follows. We describe the \textit{HST} COS observations and atmospheric models with the fitting procedure in Sections\,\ref{sec:obs}\,\&\,\ref{sec:fit}, respectively. We compare the atmospheric parameters (\Teff\ and \logg) obtained in this study with the previous spectroscopic and photometric studies along with their mass distributions in Sections\,\ref{sec:comp}\,\&\,\ref{sec:massd}, respectively. Taking advantage of the comparative study, we identify outliers comprising interesting binary candidates that exhibit large deviations from the published studies described in Section\,\ref{sec:outliers}. Finally, we discuss and conclude our study in Sections\,\ref{sec:discus}\,\&\,\ref{sec:conc}, respectively.

\section{Observations}\label{sec:obs}
Starting with \textit{HST}'s Cycle\,18, we have carried out seven COS snapshot surveys of white dwarfs. The analysis of these observations has so far largely focused on the sources displaying photospheric metal contamination from the accretion of planetary debris. An initial statistical study of 85 young DA white dwarfs ($20-200$\,Myr, $17\,000\lesssim\Teff\lesssim27\,000$\,K) reported their atmospheric parameters and found that 56\,per cent of these stars displayed traces of metals in their spectra \citep{Koester2014}. A number of individual results include the first detailed assessment of the diversity in the abundances of planetary debris \citep{boris2012}, the detection of water-rich extra-solar minor planets \citep{farihi2013, hoskin2020} as well as rocky planetary debris in two white dwarfs of the Hyades \citep{farihi2013b}. In addition, the COS snapshot spectra were used to identify absorption of molecular hydrogen in three cooler DA white dwarfs \citep{Xu2013}, and the first far-UV study of an extremely low-mass white dwarf \citep{hermes2014}. 

Before \textit{Gaia} Data Release~2 parallaxes were available, the snapshot targets of the \textit{HST} programs 12169, 12474, 13652, 14077, 15073, and 16011 were selected from the Palomar Green (PG) Survey \citep{Liebert2005} in the northern hemisphere, and ESO SN Ia Progenitor surveY (SPY) \citep{Koester2009} in the southern hemisphere, accounting for the majority ($\approx73$\,per cent) of the DA sample presented here. The remaining targets (program 16642) were drawn from the \textit{Gaia}-based white dwarf catalogue of \citet[GF21, hereafter]{nicola2021}. The main criteria of the target selection were (i) the stars had effective temperatures ranging from $12\,000\lesssim\Teff\lesssim33\,000$\,K, and (ii) had predicted fluxes $\gtrsim5\times10^{-14}~\mathrm{erg\,cm^{-2}\,s^{-1}\,\mbox{\AA}^{-1}}$ at 1300\,\AA, with the goal to achieve a signal-to-noise ratio (SNR) $\ga15$ at 1300\,\AA\ in the short ($\le2000$\,s) snapshot exposures. In addition, the latest survey (program 16642) was limited to stars within 100\,pc. Given the intrinsic selection effects of the \cite{Liebert2005} and \cite{Koester2009} samples, and the fact that not all \textit{HST} snapshot targets were observed, the COS white dwarf snapshot survey is not statistically complete, but representative of nearby warm white dwarfs. The corresponding optical magnitudes of the  observed sources is $13\lesssim G\lesssim17$, with a median of $G=15.2$.

All snapshot targets were observed using the G130M grating at the 1291\,\AA\ central wavelength, covering the wavelength range 1130--1430\,\AA, with a gap at 1278--1288\,\AA\ due to the space between the two detector segments. The exposure times of the COS observations ranged from 400 to 2000\,s, with a median of 1200\,s, and a median SNR of 25.7. Because of the limited time available in a snapshot observation, we used only two of the four available FP-POS dither settings which limited somewhat our ability to mitigate against fixed pattern noise, however, we found that it did not affect the results derived from our analysis. We have used the flux-calibrated spectra retrieved from the \textit{HST} archive that are processed with COS pipeline CALCOS (v.3.3.4). 

We report the COS spectroscopy of \totalnumber\ DA white dwarfs observed between 2010 September 17 and 2023 August 02, where we excluded stars with known non-degenerate close binary companions (the observation of the non-DA white dwarfs will be analysed elsewhere). This sample includes the first spectroscopy study of 49 white dwarfs identified by GF21.

\section{Atmospheric models and Fitting}\label{sec:fit}
We have used an updated grid of pure hydrogen atmosphere models computed with the code of \cite{Koester2010} to fit the calibrated \textit{HST} COS spectra of the DA white dwarfs. The grid includes models for $7 \leq\log g < 9.25$ in steps of 0.25\,dex and $3000 < \Teff < 80\,000$\,K. The input physics and numerical methods of the atmosphere code  are described in detail in \cite{Koester2010}. Most importantly we use the Stark broadening profiles of \citet[TB09, hereafter]{Tremblay2009}. Since 2010, numerous improvements have been added to the code (non-ideal effects in the equation of state, new atomic data, collision-induced absorption, and more), but most of these are not important in the high-temperature range of this study. The exceptions are re-calculations of the unified profiles of Ly$\alpha$ and Ly$\beta$. While the basic physical effects are described in the work of Allard and collaborators, e.g. \cite{1994Allard, Allard1999}, and numerous later papers, we have used our own improved numerical procedures and new atomic data to calculate the line profiles used in this work \citep{Santos2012, Hollands2017}. The main other difference in the updated models is that ML2$/\alpha$ convection is using a mixing-length value of $l/H_{\rm P} = 0.8$ instead of 0.6, where $H_{\rm P}$ is the pressure scale height. However, this calibration is of little relevance here, since the vast majority of objects in our catalogue used for comparison are too hot ($\Teff >13\,000$\,K) for efficient convection. For the same reason, we have neglected the effects found in detailed 3D convective simulations of \citet{Tremblay2013}: while the onset of convective instabilities happens at $\approx$18\,000\,K, convective effects on predicted fluxes only become significant below $\approx$13\,000\,K, hence it is only relevant for $\approx$1\,per cent of the objects in this work. We have used local thermodynamic equilibrium (LTE) models because non-LTE effects are only noticeable on Balmer lines for $\Teff\geq40\,000$\,K \citep{Tremblay2011}.

\begin{figure*}
\centering
\includegraphics[width=0.9\textwidth]{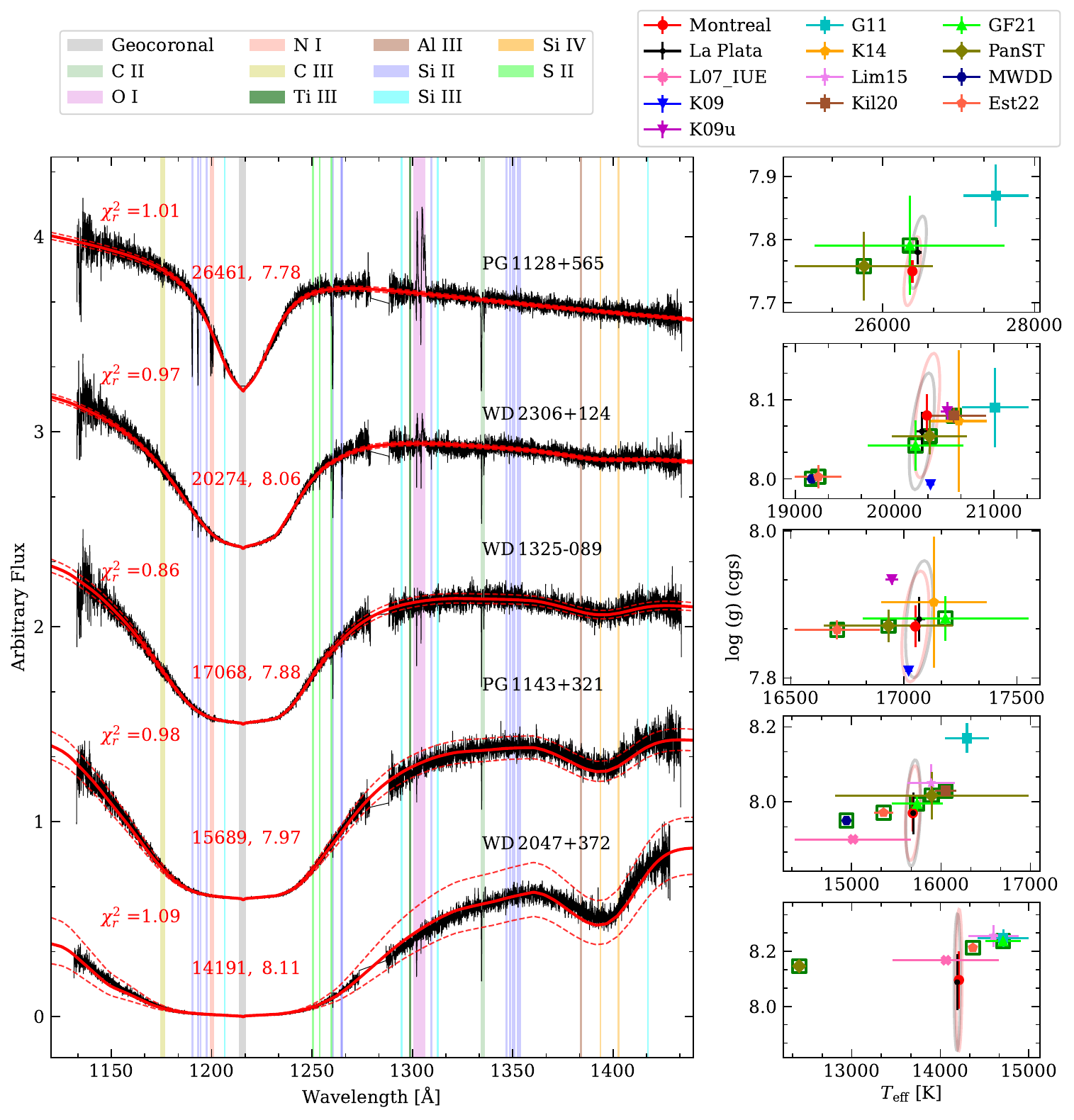}
\caption{\textit{Left}: Five examples of \textit{HST} COS UV spectra (black) of DA stars spanning the temperature range of our sample, sorted by \Teff. The spectra are normalised to their mean fluxes, and vertically offset by suitable amounts for clarity. The best-fit models to the spectra (La Plata) are shown as red solid lines with the $1\sigma$ uncertainties on the parameters indicated as red dashed lines. The best-fit \Teff\ and $\log g$ values are given by the red labels. The wavelength regions shaded by coloured bands represent the masks we adopted for ISM and photospheric absorption lines and the geocoronal emission lines (see the labels in the box above and Table\,\ref{tab:mask_lines}). \textit{Right}: Atmospheric parameters of the five stars in the \Teff\ vs \logg\ plane. Shown are the 95\,per cent confidence contours measured from the COS data (Montreal in light-red colour and La Plata in grey colour), as well as published parameters. Photometric studies are indicated by green open squares, all other symbols are derived from spectroscopic studies. The abbreviations in the legend (top right panel) are: L07\_IUE\,=\,\citet{Lajoie2007}, K09\,=\,\citet{Koester2009}, K09u as K09 but with updated models, G11\,=\,\citet{Gianninas2011}, K14\,=\,\citet{Koester2014}, Lim15\,=\,\citet{Limoges2015}, Kil20\,=\,\citet{Kilic2020}, GF21\,=\,\citet{nicola2021}, Est22\,=\,\citet{Esteban2022}, PanST\,=\, this work, and MWDD\_H\,=\,parameters from MWDD using pure-H models.}
\label{fig:spec_fit_wd_good}
\end{figure*}

To determine the atmospheric parameters \Teff\ and \logg, we fitted Ly$\alpha$ and the rest of the continuum with the model atmospheres by minimising the reduced $\chi_\mathrm{r}^{2}$ using the non-linear least-squares method known as trust region reflective algorithm (\texttt{trf}) \citep{byrd1987} of \texttt{scipy optimize}.
We masked the strong interstellar and metal lines as they will pull the fit below the true continuum level, and thus lead to inaccurate parameters. For masking the metal absorption lines, we chose a reasonable width of 0.5\,\AA\ around the central wavelength as provided in Table\,\ref{tab:mask_lines} and shown in Fig.~\ref{fig:spec_fit_wd_good}. This width corresponds to a velocity range of $\sim120$\,km\,s$^{-1}$ at 1250\,\AA,  sufficient to account for the line of sight motion and the gravitational redshift of the white dwarfs. Finally, we also masked 1213--1217\,\AA\ and 1300.5--1306.5\,\AA, which are affected by the geocoronal emission lines of Ly$\alpha$ and the \ion{O}{i} triplet.

\begin{table}
\centering
\caption{A list of lines that were masked in the analysis, along with their vacuum wavelengths. The lines can have both interstellar or photospheric contributions, except those flagged by $^{*}$ which are entirely photospheric.}
\begin{tabular}{ll}
\hline
Ion & Vacuum Wavelength [\AA]\\
\hline
\Ion{N}{i} & 1199.55, 1200.22, 1200.71\\
\Ion{C}{ii}  & 1334.53, 1335.70\\
\Ion{O}{i} & 1302.17, 1304.86, 1306.03\\
\Ion{Si}{ii} & 1190.42, 1193.29, 1260.42, 1304.37, 1309.45$^{*}$\\
\Ion{S}{ii} &  1250.58, 1253.80, 1259.52\\
\Ion{Si}{iii} & 1206.51, 1294.54$^{*}$, 1296.72$^{*}$,1298.89$^{*}$,\\
& 1312.59$^{*}$, 1417.24$^{*}$\\
\Ion{Si}{iv} & 1393.75, 1402.77\\
\Ion{C}{iii} & 1174.93$^{*}$, 1176.37$^{*}$\\
\Ion{Al}{iii} & 1384.13$^{*}$\\
\ion{Ti}{iii} & 1298.99$^{*}$\\
\hline
\label{tab:mask_lines}
\end{tabular}
\end{table}

For fitting the spectra, the observed fluxes ($F_{\lambda}$) were compared with the model Eddington fluxes ($H_{\lambda}$) using the following relation:
\begin{equation}
F_{\lambda}= 4\pi(R/D)^{2} H_{\lambda} (\Teff,\logg)
\label{eqn:fit}
\end{equation}
{\noindent}where, \Teff\, \logg\, and, parallax (hence, $D$ which is the distance to the Earth) are considered free parameters of the model. While performing the fit with \texttt{trf} method, the bounds were specified in the free parameters where the bounds for the \Teff\ and \logg\ correspond to the model grid limits of mass-radius (M-R) relations, while the distances are constrained using the \textit{Gaia} DR3 parallaxes ($\varpi$) and its errors taken from the white dwarf catalogue of GF21. In Eqn.\ref{eqn:fit}, $R$ is the radius obtained from the M-R relation corresponding to the best fit \Teff, \logg, and, $D$ from \textit{Gaia} parallax. The model fluxes were reddened using the Fitzpatrick extinction law \citep{Fitz1990, Fitz1999} in the \texttt{extinction}\footnote{\url{https://extinction.readthedocs.io/en/latest/index.html}} code. The extinction values are considered from GF21 that were derived using 3D extinction map STILISM/EXPLORE \citep{lallement2019}. Finally, the statistical uncertainties in the fitted parameters are obtained directly from the covariance matrix of the fitting algorithm scaled by $\rm{\chi^{2}_{r}}$ to account for the goodness of fit.

We implemented two different M-R relations in our fitting routine to obtain the radius and mass by interpolating the \Teff\ and \logg\ model grids of DA white dwarfs. The two models used for M-R relation are the one from the Montreal\footnote{\url{http://www.astro.umontreal.ca/~bergeron/CoolingModels/}} which uses theoretical evolutionary sequences of \cite{Bedard2020} corresponding to thick H layers, and the one from La Plata \citep{Althaus2013, Camisassa2016, Camisassa2019} which uses the model grid of DA generated from the {\tt LPCODE} \citep{Althaus2005} stellar evolutionary code\footnote{\url{http://evolgroup.fcaglp.unlp.edu.ar/TRACKS/tracks.html}}. The details of the model parameters are provided in Table\,\ref{tab:M-Rrel}. Both models are appropriate for a progenitor metallicity of $Z = 0.02$. The main differences to be noticed between the models are the assumption of core compositions and the thickness of H layers for different white dwarf masses. La Plata models are more appropriate specifically for low-mass stars  ($<0.4\,\Msun$) which assume a He core and a thicker H envelope ($\simeq10^{-3}\,\rm{M_{H}}/M_{WD}$). The Montreal and La Plata sequences have similar cooling ages for mass $\simeq0.6$\,\Msun\ ($\logg\simeq8$), but differ vastly for lower ($<0.4\,\Msun$) and higher masses ($>1.0\,\Msun$). In the following, we will refer to the two different M-R relations simply as ``Montreal'' and ``La Plata''.

\begin{figure}
\centering
\includegraphics[width=\columnwidth]{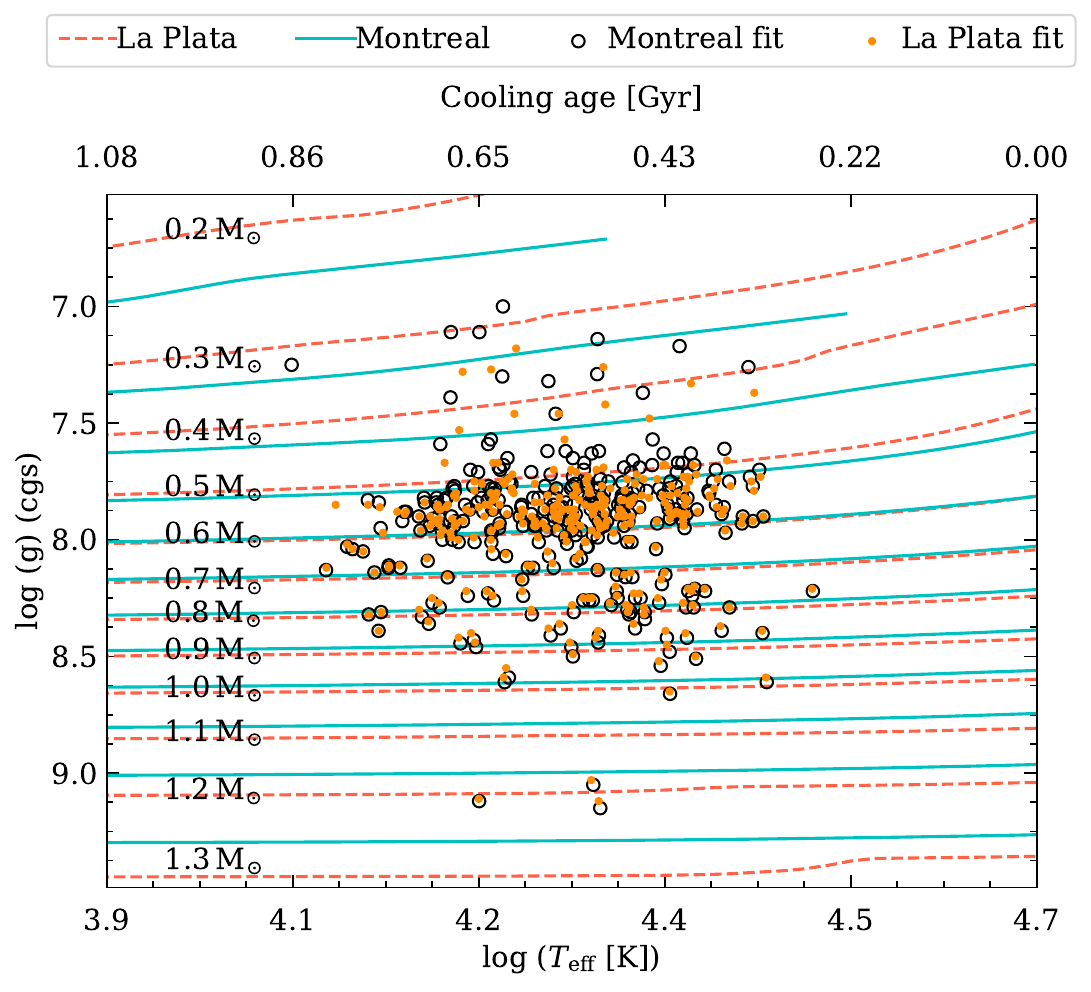}
\caption{The \Teff\ and \logg\ values for \totalnumber\ white dwarfs derived from $\chi^{2}$ fitting of the COS spectra (Montreal: black open circles; La Plata: orange dots). The cooling age in Gyr is shown in the top axis for \logg = 8. The solid cyan and red dashed lines represent the evolutionary sequences from Montreal models with thick H envelopes, and La Plata models for spectral type DA respectively. Model masses ($0.2-1.3$\,\Msun) are labelled in the figure.}
\label{fig:teff_logg}
\end{figure}

The best-fit parameters along with the two model grids are shown in the \Teff-\logg\ plane in Fig.\,\ref{fig:teff_logg}. The models match for stellar mass of 0.6\,\Msun\ as they consider the same value for the H envelope ($\simeq10^{-4}\,\rm{M_{H}}/M_{WD}$). The difference between the model grids increases in the low mass ($<0.5\,\Msun$) and high mass end ($\geq1.0\,\Msun$) where 20\,per cent (Montreal) and 10\,per cent (La Plata) of the targets in our sample are located. This difference is due to the consideration of different core compositions and thicknesses of the H layers in their models. As the mass of the H envelope decreases, the \logg\ increases for a given mass and \Teff\ of the white dwarf. This is pointed out by \cite{Romero2019} who showed that not accounting for the dependence of H envelopes on the models can result in an overestimate of the stellar mass. 

The fit parameters of 49 white dwarfs with no previously reported spectroscopic measurements in literature are provided in Table~\ref{tab:phot_params_wd}. A full catalogue with the atmospheric parameters is made available online through Vizier. 

\begin{table}
\centering
\caption{Model parameters of the two mass-radius relations from the Montreal and La Plata models for a progenitor metallicity of $Z=0.02$. }
\addtolength{\tabcolsep}{-3pt}
\begin{tabular}{lll}
\hline
Parameters	&	Montreal$^{1}$	&	La Plata\\
\hline
\Teff	&	$1460-150\,000$\,K	&	$2750-80\,000$\,K\\
\logg 	&	$6.7-9.3$	&	$6-9.45$\\
Mass ($\rm{M_{WD}}$)	&	$0.2-1.3$\,\Msun	&	$0.2-1.3$\,\Msun\\
Core composition	&	CO core 	&	He Core $(\rm{M_{WD}}<0.5\,\Msun)^{2}$\\
	& entire mass range		&	CO core $(0.5 \leq \rm{M_{WD}} \leq 1.0 \,\Msun)^{3}$\\
	&		&	O-Ne core $(\rm{M_{WD}}\geq 1.1\,\Msun)^{4}$\\
H envelope mass 	&	$\sim10^{-4}$ 	&	$\sim10^{-3}$ ($\rm{M_{WD}} \leq 0.32\,\Msun)^{2}$\\
($\rm{M_{H}}/M_{WD}$)	&	entire mass range 	&	$\sim10^{-3.5}-10^{-4.5}$ \\
&&$(0.5 \leq \rm{M_{WD}} \leq 0.88\,\Msun)^{3}$\\
	&	 	& $\sim10^{-6} (\rm{M_{WD}}\geq 1.1\,\Msun)^{4}$\\
\hline
\label{tab:M-Rrel}
\end{tabular}
\footnotesize\flushleft
$^{1}$\cite{Bedard2020}, $^{2}$\cite{Althaus2013}, $^{3}$\cite{Camisassa2016}, $^{4}$\cite{Camisassa2019}
\end{table}

\subsection{Atmospheric parameters}\label{sec:params}
To illustrate the results from our fitting procedure, we show the best-fit models (using the La Plata M-R relation) superimposed on the COS spectra for five white dwarfs spanning the full range in temperature covered by the snapshot surveys in the left panel of Fig.\,\ref{fig:spec_fit_wd_good}. Overall, the \Teff\ and \logg\ derived from the COS data agree reasonably well with the published results (right panels). 

As the atmospheric parameters \Teff\ and \logg\ are highly correlated in the fit, we calculated the 95\,per cent confidence ellipse from the eigenvalues and eigenvectors of the covariance matrix, which is shown for Montreal (light red) and La Plata (grey) fits in the right panels of Fig.\,\ref{fig:spec_fit_wd_good}. The typical statistical uncertainties in \logg\ are 0.02\,dex which increases to 0.04\,dex if we consider the uncertainties in parallaxes, while the uncertainties in \Teff\ that are typically $\sim$50\,K remain unchanged. For stars with $\Teff<20\,000$\,K, we note that the broad Ly$\alpha$ satellite \ion{H}{$^{+}_{2}$} feature appears at 1380--1410\,\AA\ \citep{1985A&A...142L...5K}, which increases in strength for decreasing temperatures. We find that this feature is overall well-fitted by the models.

We collected the published values for \Teff\ and \logg\  for the stars in our sample available from the Montreal White Dwarf Database \citep[MWDD]{mwdd}\footnote{\url{https://www.montrealwhitedwarfdatabase.org/}}, and show these parameters and their $1\sigma$ uncertainties in the right-hand-side panels of Fig.\,\ref{fig:spec_fit_wd_good}. We also include our own fit to where the olive diamond denotes the value derived from Pan-STARRS photometry.
We find good fits ($\chi^{2}_\mathrm{r}\simeq1$) for most stars in our sample, and the atmospheric parameters of these stars (Fig.\,\ref{fig:spec_fit_wd_good}) typically agree with literature values within the uncertainties ($3\sigma$). However, we were unable to obtain a reasonable fit for a small fraction of stars which are further discussed in detail in Section\,\ref{sec:outliers}. 

We note that most published analyses are based on optical spectroscopy and photometry, and some studies are likely using the same observations or even parameters from earlier papers \citep[e.g.][]{Liebert2005, Gianninas2011, Limoges2015}. These atmospheric parameters were derived over several decades using a variety of techniques and models, some of which relied on free parameters to account for non-ideal gas effects (prior to TB09). Hence, the spread in literature values should not be taken as a realistic representation of atmospheric parameter uncertainties. 
\begin{table*}
\centering
\caption{Previous studies used for comparative analysis with our work}
\addtolength{\tabcolsep}{-7pt}
\begin{tabular}{cccccc}
\hline
\hline
Study	&	Sample \& Data	&	Wavelength (\AA)	&	Models	&	Methods	& common stars\\\hline
\noalign{\smallskip}
\multicolumn{6}{c}{Spectroscopy} \\
\noalign{\smallskip}
\hline
\cite{Liebert2005}	&	PG survey (348 WDs), optical spectra 	&	3500$-$6000	& \cite{Liebert2005} &	Balmer line fitting (normalisation)	&	51	\\
\cite{Koester2009}	&	SPY (615 WDs), high-resolution &	3500$-$6650	& 1) VCS profiles &	Balmer line fitting (normalisation)	&	123	\\
& optical spectra (UVES)& &2) This work (TB09) &&\\
\cite{Gianninas2011}	& optical spectra (1100 WDs) 	&	3500$-$6000	& TB09	& Balmer line fitting (normalisation)	&	196	\\\hline
%
\noalign{\smallskip}
\multicolumn{6}{c}{Spectrophotometry}\\
\noalign{\smallskip}
\hline
This work	&	\textit{HST} COS (307 WDs) &	1130$-$1435	& This work	&	$\chi^{2}$ fitting, no normalisation, \textit{Gaia} parallaxes, & -\\
&&&& extinction, two M-R relations &		\\
\cite{Lajoie2007}	& \textit{IUE} spectra (140 WDs) &	1150$-$3150	& \cite{Liebert2005}	& free parameter (\Teff), \logg\ fixed to optical, &	15	\\
&&&& distance from two methods &\\
\cite{Koester2014}	&	\textit{HST} COS (85 WDs)  &	1130$-$1435	& \cite{Koester2010}		&	free parameter (\Teff), no parallaxes, & 84\\
&&&&no extinction, \logg\ fixed to optical &\\\hline
\noalign{\smallskip}
\multicolumn{6}{c}{Photometry}\\
\noalign{\smallskip}
\hline
\cite{Kilic2020} &	SDSS($u$)+Pan-STARRS ($grizy$)	&	3500$-$9300	&	TB09	&	photometric technique, no extinction	&	66	\\
&&&&\textit{Gaia} parallaxes &		\\
MWDD \citep{mwdd}	&	Pan-STARRS ($grizy$) &	3500$-$9300	& TB09	&	photometric technique, \textit{Gaia} parallaxes &	188	\\
\cite{nicola2021} &	\textit{Gaia} EDR3 ($G$, $G\rm{_{BP}}$, $G\rm{_{RP}}$)	&	3500$-$9300	& TB09	& photometric technique, \textit{Gaia} parallaxes & 309\\
\cite{Esteban2022} &	\textit{Gaia} DR3 (JPAS) &	3500$-$9300	& Koester models & photometric technique, \textit{Gaia} parallaxes & 225\\
&	&	& TB09 profiles	& La Plata M-R  & \\
This work	&	Pan-STARRS ($grizy$)	&	3500$-$9300	&	TB09	&	same as GF21	&	257\\
\hline
\label{tab:compare_table}
\end{tabular}
\end{table*}

\section{Comparison of the COS atmospheric parameters with previous studies}\label{sec:comp}

In the following sections, we compare the atmospheric parameters derived from fitting the COS spectroscopy with the published spectroscopic and photometric studies, which we selected from the available literature to have sufficient overlap in targets with our snapshot sample (Table\,\ref{tab:compare_table}).  

\subsection{Comparison with spectroscopic studies}

\begin{figure*}
\centering
\includegraphics[width=\textwidth]{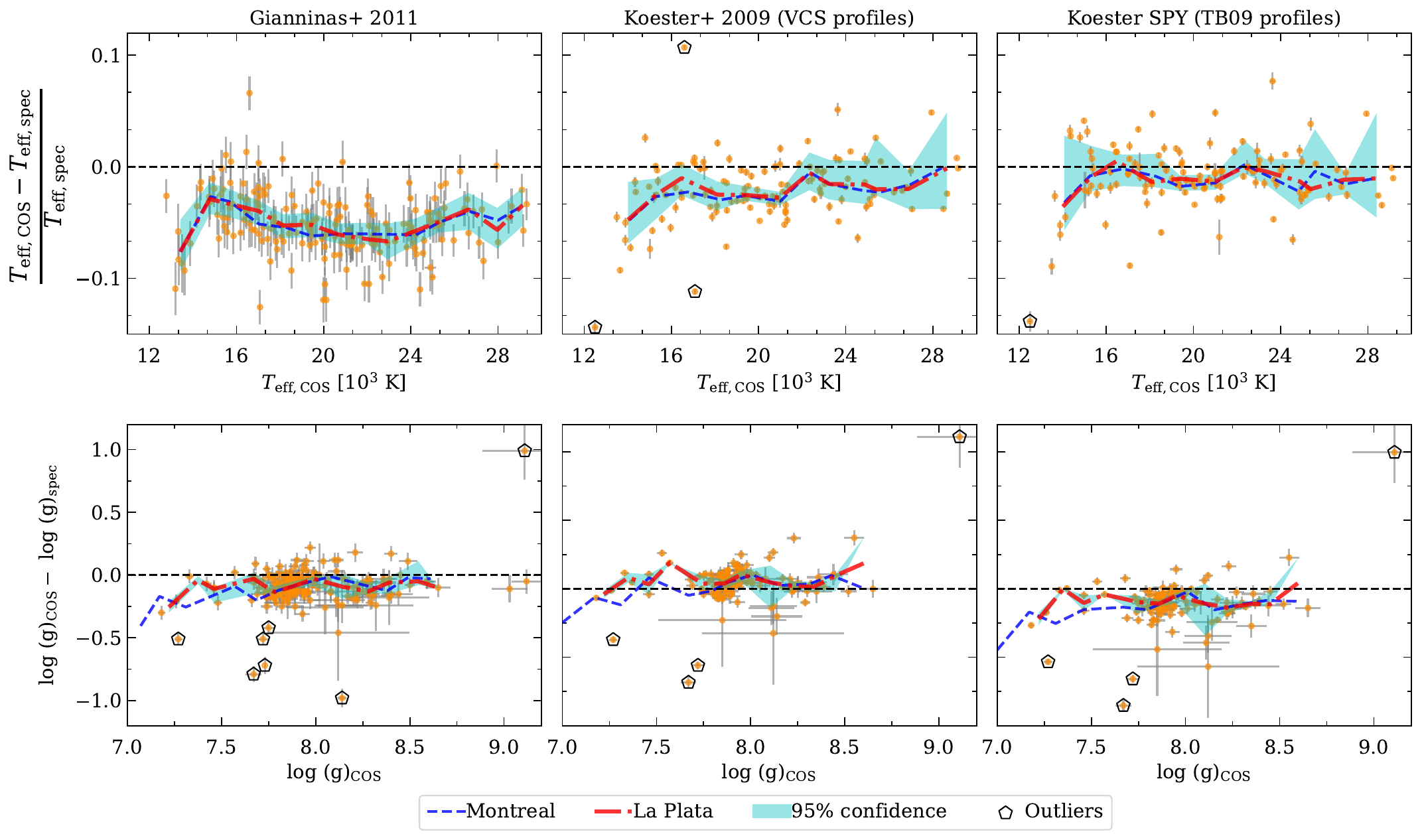}
\caption{Top panels: Differences in \Teff\ measured from the COS spectra ($T_{\rm{eff,COS}}$) and  \Teff\ from optical spectroscopic studies ($T_\mathrm{eff,spec}$), normalised to  $T_{\rm{eff,spec}}$, as a function of  $T_{\rm{eff,COS}}$ for $T_\mathrm{eff,spec}$ from  G11 (left), K09 (VCS Stark profiles; middle), and K09 but with the updated Stark profiles from TB09 (right). Bottom panels: same as the top panels but for \logg.  The dashed (blue, Montreal) and dash-dotted (red, La Plata) lines illustrate the median values with a non-uniform binning in steps of $\approx1000$-1500\,K for \Teff\ and $\approx$0.1-0.15\,dex for \logg. The shaded cyan colour denotes the 95\,per cent confidence interval for the corresponding median values obtained by boot-strapping. The outliers (Sect.\,\ref{sec:outliers}) are marked by black pentagons.}
\label{fig:spec_uv_comp}
\end{figure*}

\subsubsection{Comparison with optical spectroscopic studies}
We identified four optical spectroscopic studies that have a sufficiently large overlap in targets with our COS sample to warrant a comparison (Table\,\ref{tab:compare_table}). The parameters determined in these studies are based on the traditional technique of fitting the synthetic spectra to the normalised Balmer lines with the continuum set to unity using the non-linear least-squares method. 

\citet[G11; hereafter]{Gianninas2011} provided atmospheric parameters of 1100 DA white dwarf stars by analysing the optical spectra ($\approx3500-6000$\,\AA) obtained from several different telescopes. They used the model atmospheres as described in \cite{Liebert2005, Tremblay2011} with improved Stark broadening profiles of TB09. We found 194 stars in common with their catalogue which is the largest overlap with any optical spectroscopic study. 

Similarly, \citet[K09; hereafter]{Koester2009} carried out a high-resolution optical spectroscopic study of 615 DAs. Their model atmospheres were based on older grids of VCS Stark profiles \citep{vidal1973} and did not include the improved hydrogen Stark broadening profiles of TB09, when compared with the updated models used in this work. In order to compare our COS results like-for-like, we re-fitted the 123 common stars following the same method as described by K09, but using updated models. The main difference to results in K09 is a systematically higher \logg, which is mostly due to the use of updated Stark broadening profiles. 

The differences in \Teff\ and \logg\ between our COS results and the optical studies are shown in Fig.\,\ref{fig:spec_uv_comp}.  We note that the G11 \Teff\ and \logg\ values are, on average, systematically higher by $\simeq5$\,per cent and 0.1\,dex, respectively, than those derived from the COS spectroscopy. The COS \Teff\ estimates also show a systematic negative offset of three\,per cent compared to K09 (who used VCS profiles). However, comparing to the re-fitted K09 parameters using the updated models, this offset reduces to 1.5\,per cent, bringing the UV values being in closer agreement. Comparing the \logg\ measurements, we find that the COS results are 0.1\,dex higher than the original K09 values, while $0.1-0.15$\,dex lower when compared to the K09 re-analysis using updated models.  

\begin{figure}
\centering
\includegraphics[width=\columnwidth]{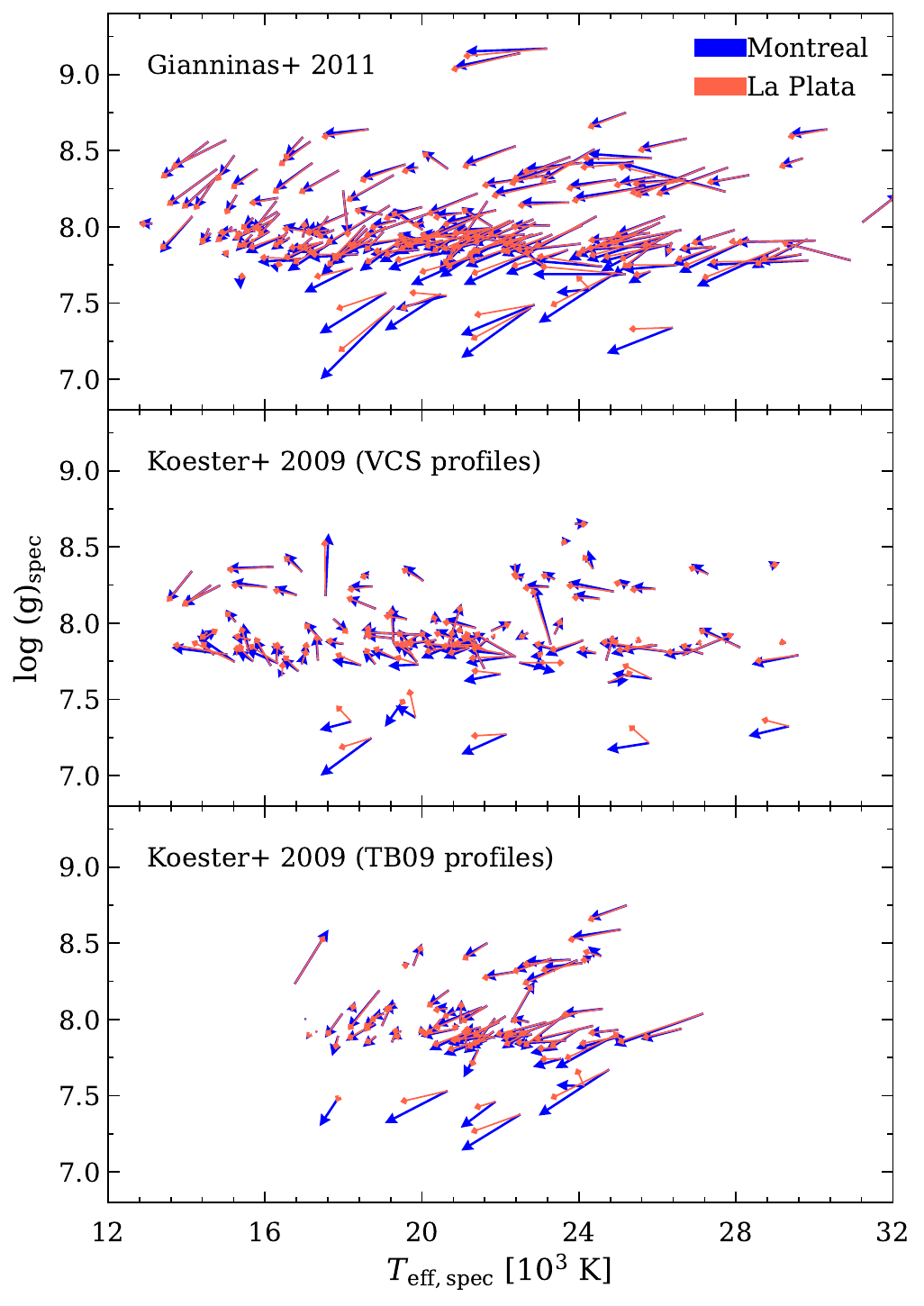}
\caption{Comparison of the COS atmospheric parameters with those derived from optical spectroscopy (G11: top; K09 VCS Stark profiles: middle; K09 with the updated models using TB09 profiles: bottom) in the \Teff-\logg\ plane. The blue and red arrows are the parameters from the Montreal and La Plata M-R fits, where the beginning of the arrow indicates the optical parameters, and the arrowhead those derived from the COS data. The sources with \Teff\ difference more than 10~per cent and \logg\ difference larger than 0.3\,dex have been excluded to avoid crowding.} 
\label{fig:spec_uv_arrow}
\end{figure}

The differences between the atmospheric parameters derived from optical data and from the COS observations are shown in the \Teff\--\logg\ plane in Fig.\,\ref{fig:spec_uv_arrow} to illustrate the correlations between the two parameters. It is clearly apparent that the COS \Teff\ and \logg\ are systematically offset towards lower values compared to G11, while there is more scatter in the comparison with K09. The differences between the Montreal and La Plata fit parameters are noticeable for $\logg<7.8$, corresponding to masses $<0.5\,\Msun$, as expected due to different H envelope masses and core compositions in the models, as discussed earlier. 

\subsubsection{Comparison with UV spectroscopic studies}
\cite{Lajoie2007} presented a comparative study of \Teff\ for 140 DA white dwarfs determined from optical (Balmer lines) and UV spectra covering the wavelength regions $1150-1970$\,\AA\ obtained with the short-wavelength primary camera (SWP) onboard \textit{IUE} and $1850-3150$\,\AA\ using the long-wavelength primary (LWP) and redundant (LWR) cameras. For the UV fits, they fixed \logg\ to the values derived from the optical spectra, and estimated distances using $V$-band magnitudes and a distance modulus derived from the scaling factor of the models. Our COS analysis differs both in  wavelength coverage and methodology, as we are determining \logg\ from the flux-calibrated COS spectra and the \textit{Gaia} parallaxes. Comparing their results with COS, we note that the COS \Teff\ of 15 common stars are higher by two\,per cent than \cite{Lajoie2007}, while the \logg\ values are on average lower by 0.03\,dex. 

\begin{figure}
\centering
\includegraphics[width=0.9\columnwidth]{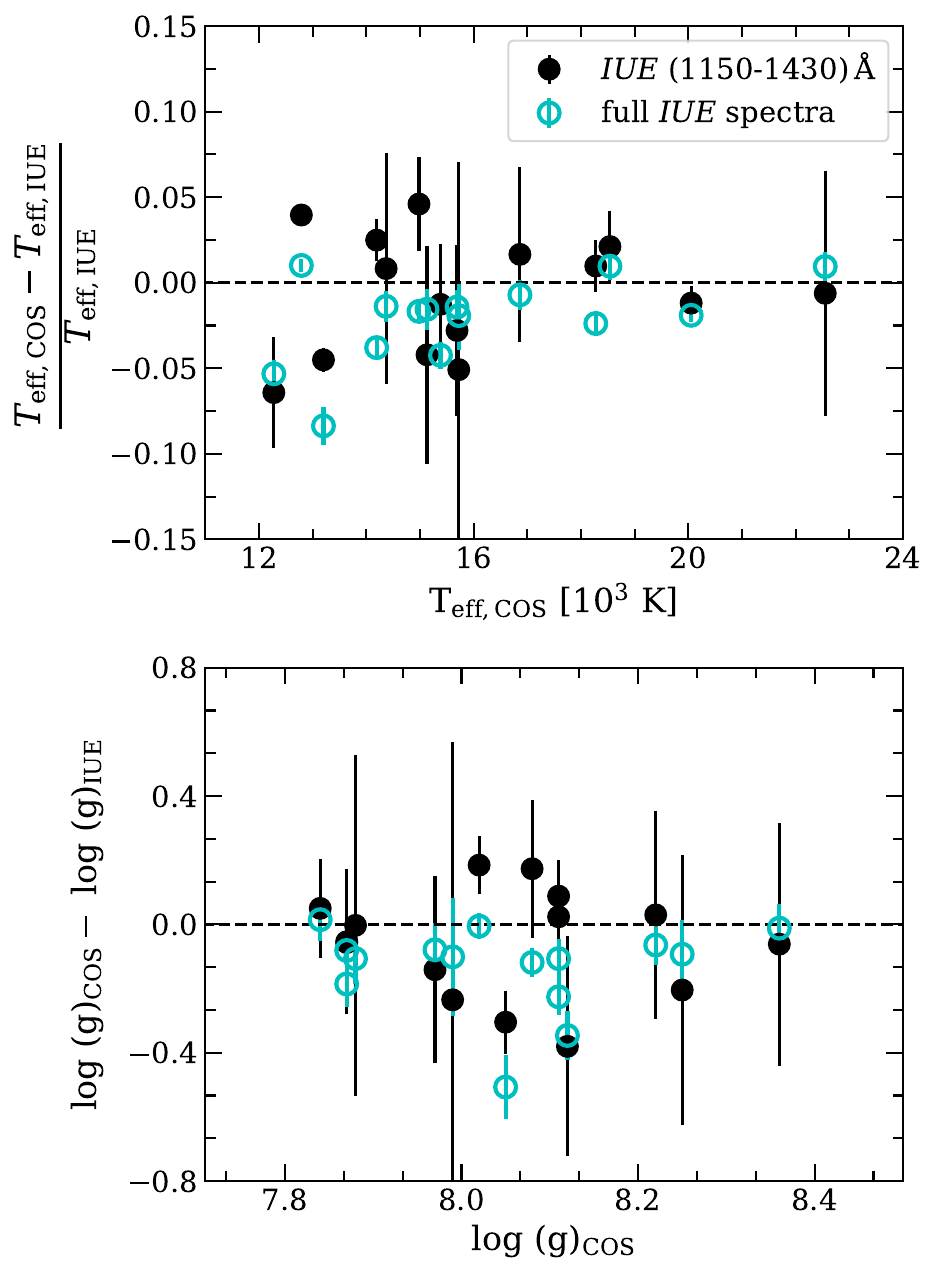}
\caption{Differences in \Teff\ (top) and \logg\ (bottom) between the COS and \textit{IUE} measurements. The black dots and cyan circles denote the parameters obtained considering $1150-1430$\,\AA\ and the entire wavelength range ($1150-1970$\,\AA) of \textit{IUE} spectra, respectively.}
\label{fig:iue_cos_comp}
\end{figure}

To identify whether the difference is due to the updated models or data, we fitted the \textit{IUE} far-UV spectra with the same models and fitting procedure as in our COS analysis. As there is a wavelength overlap of the \textit{IUE} data from the short wavelength prime (SWP) camera with COS, we derived the parameters for two cases, first considering a similar spectral region as COS ($1150-1430$\,\AA) and second using the entire spectral coverage. We find a scatter of five\,per cent in the \Teff\ differences for stars having $\Teff <16\,000$\,K with the COS \Teff\ being lower in the latter case (see Fig.\,\ref{fig:iue_cos_comp}). Additionally, COS \logg\ are systematically lower by $\approx0.25$\,dex than the values obtained from the entire spectrum fitting of \textit{IUE}. Since the sample of common stars available for comparison is very small and the statistical uncertainties in the \textit{IUE} measurements are larger than those from our COS analysis, it is difficult to provide a definitive conclusion on the systematics present.

\begin{figure}
\centering
\includegraphics[width=0.9\columnwidth]{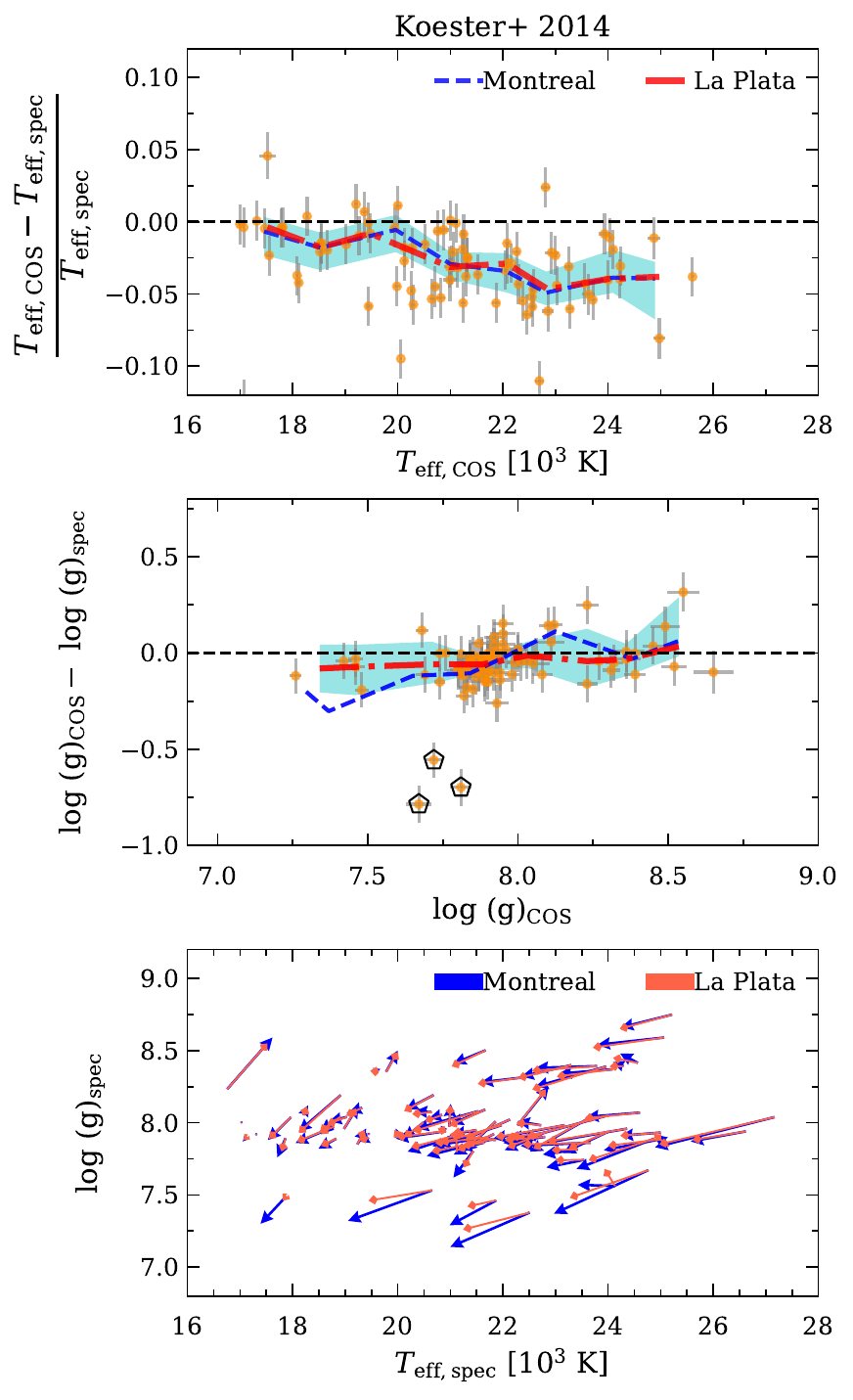}
\caption{Same as Fig.\,\ref{fig:spec_uv_comp} but for comparisons of the \Teff\ (top), and \logg\ (middle) we measured from the COS data, and the parameters derived by \citet{Koester2014} using the same data, but a different methodology (as their work pre-dated the \textit{Gaia} parallaxes). The comparison in the \Teff-\logg\ plane is shown in the bottom panel, see Fig.\,\ref{fig:spec_uv_arrow} for a description of symbols.}
\label{fig:spec_uv_comp_k14}
\end{figure}

\citet[hereafter, K14]{Koester2014} derived the parameters of 85 DA white dwarfs using the same model atmospheres and \textit{HST} COS data as that utilised in our work. However, in the absence of accurate distance and reddening measurements, they adapted a different fitting method compared to our analysis: the \logg\ values were fixed to results from optical studies as the COS spectra mainly sample the red wing Ly$\alpha$, which is insufficient to independently determine \Teff\ and \logg. Consequently, only \Teff\ was varied to obtain the best fits. The differences in \Teff\ and \logg\ between our work and that of K14 are shown in the top and middle panels of Fig.\,\ref{fig:spec_uv_comp_k14}. We note that there is an offset in the temperatures with our \Teff\ being lower than those from K14 which reaches $\simeq5$\,per cent at \Teff\ $>20\,000$\,K. This trend towards lower \Teff\ in our study is clearly evident in \Teff-\logg\ plot shown in the bottom panel of Fig.\,\ref{fig:spec_uv_comp_k14}, where we notice larger \Teff\ differences for stars hotter than 20\,000\,K. Our \logg\ measurements agree with those of K14 with 95\,per cent confidence, only for $\logg\lesssim8.0$ we notice a systematic offset, with our values being lower than those of K14. 

\subsection{Comparison with photometric studies}

\begin{figure*}
\centering
\includegraphics[width=\textwidth]{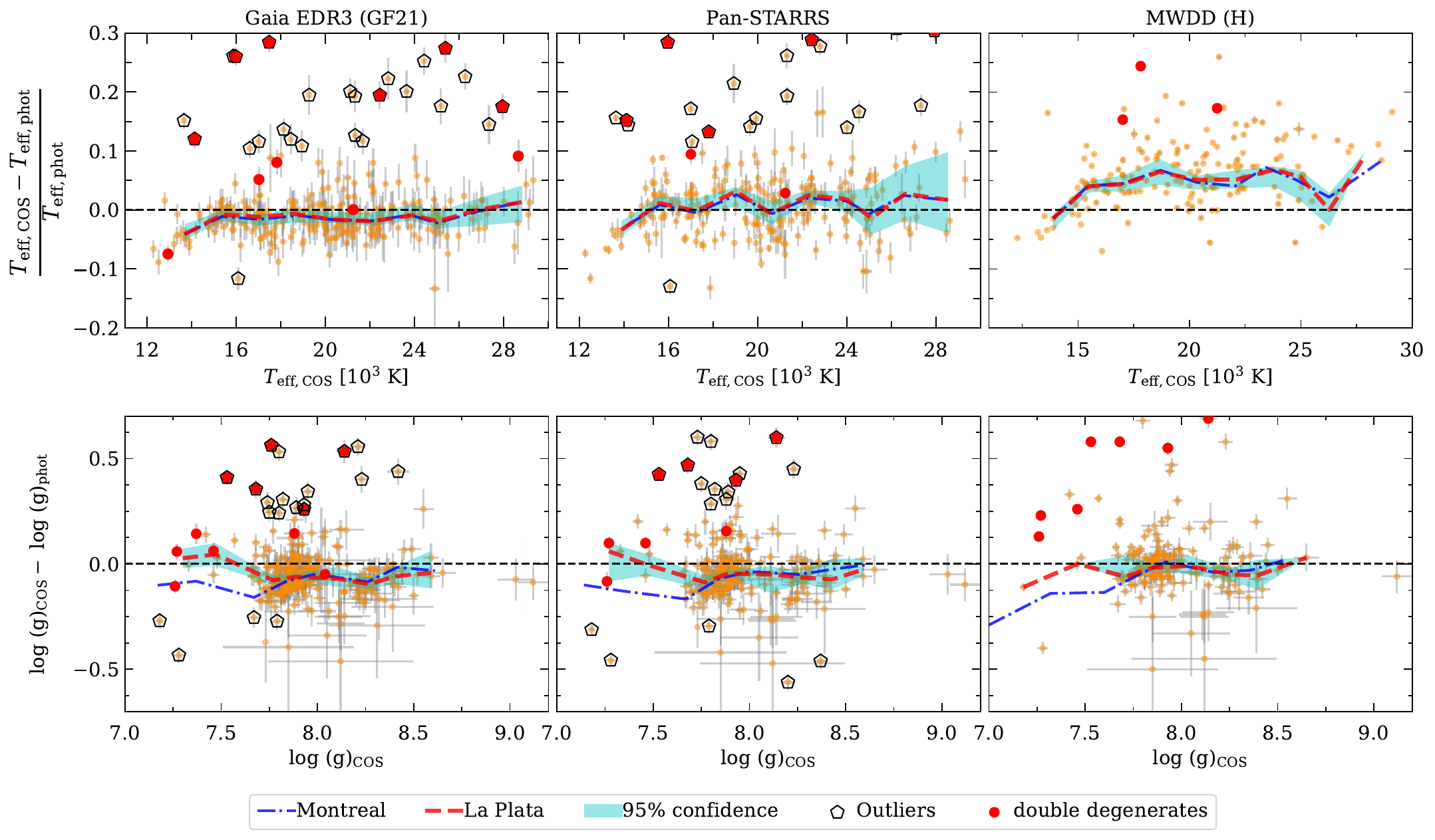}
\caption{Top panels: Differences between \Teff\ derived from the COS UV spectra ($T_{\rm{eff,UV}}$) and \Teff\ based on optical photometric studies ($T_{\rm{eff,phot}}$), normalised to  $T_{\rm{eff,phot}}$, as a function of  $T_{\rm{eff,UV}}$ for $T_\mathrm{eff,phot}$ from \textit{Gaia}~EDR3 (\citealt{nicola2021}, left), Pan-STARRS (middle) and MWDD (right). Bottom panels: same as the top panels but for differences in \logg. The photometric outliers are highlighted as black pentagons, and known double-degenerates as red dots \citep{Koester2009}, see Sect.\,\ref{sec:outliers} for more details on symbols and colours.}
\label{fig:phot_uv_comp}
\end{figure*}

\subsubsection{Comparison with \textit{Gaia} EDR3}
GF21 derived the parameters of the \textit{Gaia} white dwarf sample by fitting the \textit{Gaia} EDR3 ($G$, $G{\rm_{BP}}$, and $G{\rm_{RP}}$) absolute fluxes using three different sets of model atmospheres: pure H, pure He and mixed (H/He) compositions. GF21 used the model grid of \cite{Tremblay2011} with Ly$\alpha$ opacity of \cite{Kowalski2006} for pure-H composition, and cooling sequences of \cite{Bedard2020} for calculating the masses and radii of the white dwarfs with $M>0.46\,\Msun$, whereas, He-core models of \cite{Serenelli2001} (La Plata group) were used for lower masses. We selected the photometric estimates based on pure-H model atmospheres, appropriate for DA white dwarfs, to compare with the parameters we derived from the COS data. The differences between the \Teff\ and \logg\ values from our COS analysis and those from GF21 using the two different M-R relations were calculated. The comparisons are shown in the left panels of Fig.\,\ref{fig:phot_uv_comp}. Even though we find systematic offsets of $\simeq-1.5$\,per cent and $-0.07$\,dex in \Teff\ and \logg, respectively, the parameters agree with each other with 95\,per cent confidence. While calculating the median values and confidence levels, we have excluded few stars ($\approx10$\,per cent; see Fig.\,\ref{fig:phot_uv_comp}) that are flagged as photometric outliers. The selection criteria and additional details on these outliers are described in detail in Sect\,\ref{sec:outliers}.

\begin{figure}
\centering
\includegraphics[width=0.9\columnwidth]{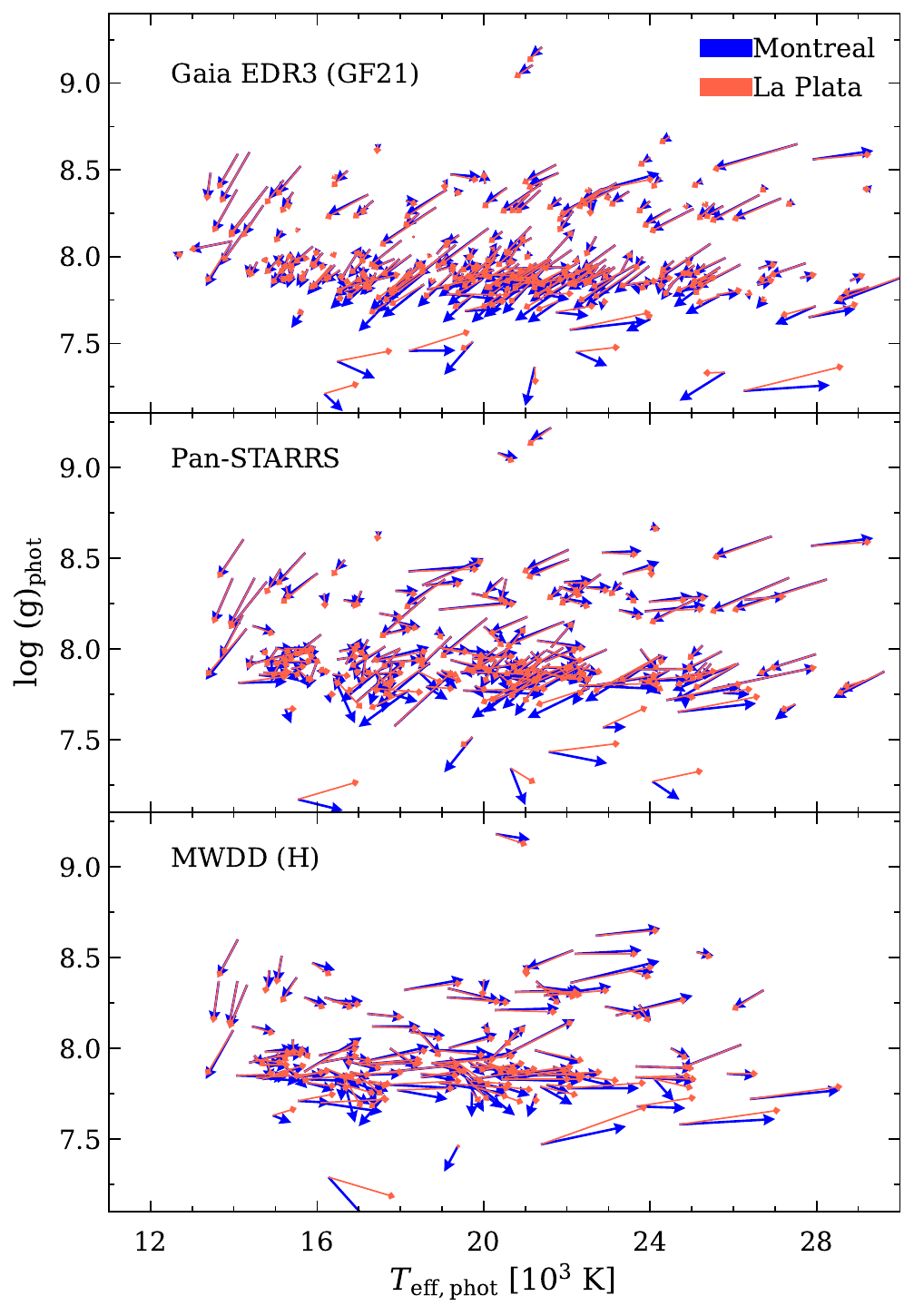}
\caption{Same as in Fig.\,\ref{fig:spec_uv_arrow} but for the comparisons of our COS results with with those based on \textit{Gaia} EDR3 (GF21,  top), Pan-STARRS (middle, see Sect.\,\ref{sec:panstarrs}), and MWDD (bottom). All photometric parameters assume pure-H model atmospheres.}
\label{fig:phot_uv_arrow}
\end{figure}

\subsubsection{Comparison with Pan-STARRS}
\label{sec:panstarrs}
MWDD \citep{mwdd} provides the basic parameters (\Teff\ and \logg) of the white dwarfs derived from the SED fitting of Pan-STARRS photometry with pure-H atmosphere models. However, they do not provide uncertainties in the estimated values. Therefore, we performed fits to the five Pan-STARRS band-passes ($grizy$) for the 257 white dwarfs in our sample that fall within the Pan-STARRS footprint. We used the same atmospheric models and methods as GF21, and we used, in addition to the photometry, the reddening and parallax values from GF21. The comparisons of photometric \Teff\ and \logg\ from Pan-STARRS (this work) and MWDD with the COS results are shown in the middle and right panels of Fig.\,\ref{fig:phot_uv_comp}. We note that the \Teff\ values that we derived using Pan-STARRS data agree well with the COS estimates as shown in the binned medians with 95\,per cent confidence (middle panel of Fig.\,\ref{fig:phot_uv_comp}). The \logg\ differences show a $-0.06$\,dex systematic offset, similar to the offset found in comparison with GF21. Comparing with MWDD parameters, we find a 5--7\,per cent offset in the \Teff\ determinations (top right panel of Fig.\,\ref{fig:phot_uv_comp}) with COS values being comparatively higher for stars hotter than 15\,000\,K, whereas the \logg\ values agree with each other.

Similar to the spectroscopic comparisons, to investigate further the systematic offsets of \Teff\ and \logg\ for photometric studies, we show trends in the \Teff--\logg\ plane as arrow plots in Fig.\,\ref{fig:phot_uv_arrow}. In comparison with the \textit{Gaia} and Pan-STARRS parameters (top and middle panels of Fig.\,\ref{fig:phot_uv_arrow}, respectively), we find that arrows for stars with $\log g>7.5$ systematically point towards the lower left, illustrating that the COS analysis results in lower \Teff\ and \logg. However, in  the comparison with the MWDD parameters (bottom panel of Fig.\,\ref{fig:phot_uv_arrow}), the arrow points preferentially to the right, indicating higher values of \Teff\ determined from the COS data for $\Teff\geq15\,000$\,K, which is contradictory to what we observe in the comparison with \textit{Gaia} and Pan-STARRS results.

Based on the comparison of parameters derived from Pan-STARRS (this work) and COS, 11\,per cent of the objects in our sample show large deviations. We found 6.5\,per cent outliers in common with those selected from \textit{Gaia}. One of the main reasons for the outliers could be the Pan-STARRS saturation in brighter magnitudes. Hence, we only consider the outliers from the comparison with \textit{Gaia} for further discussion in Sect.\,\ref{sec:outliers}.

\section{Mass distribution}\label{sec:massd}
\begin{figure}
\centering
\includegraphics[width=0.95\columnwidth]{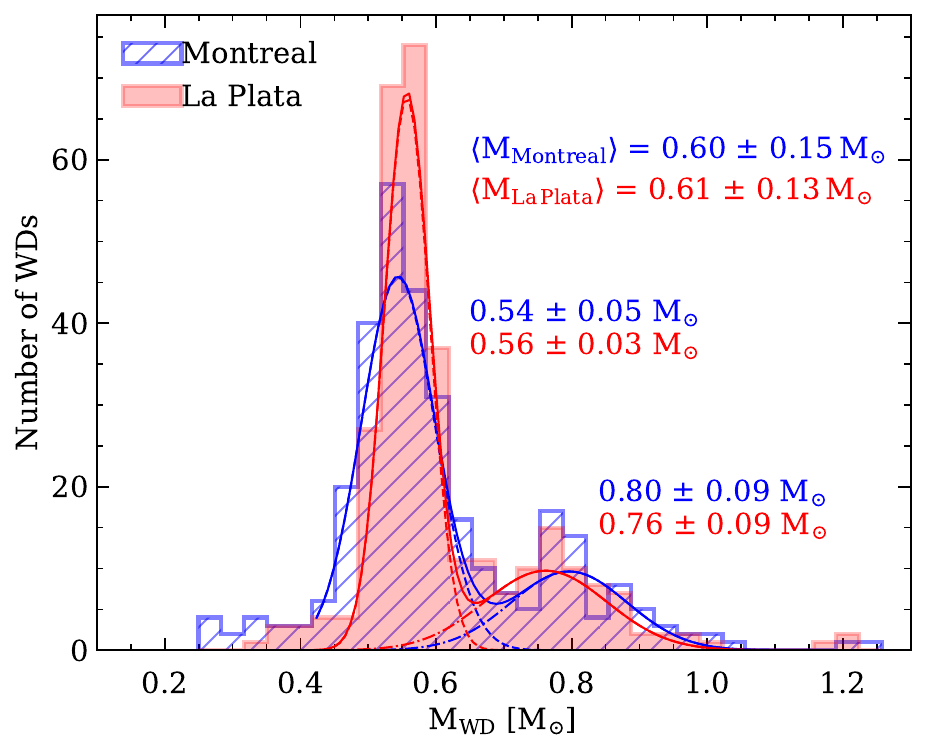}
\caption{Mass distribution of white dwarfs derived from the COS atmospheric parameters based on the fits using the Montreal M-R (\citealt{Bedard2020}, blue) and the La Plata M-R (\citealt{Althaus2013, Camisassa2016, Camisassa2019}, red). Double Gaussian fits are shown as blue and red solid lines for Montreal and La Plata fits (with individual components in dashed and dotted lines) respectively. The fit parameters for the two Gaussian components, $\mu$ and $\sigma$, are also reported, along with the mean values.} 
\label{fig:mass_wd}
\end{figure}

The mass of the white dwarfs in our sample is derived from the fitted parameters \Teff\ and \logg\ using the two M-R relations. Their distribution estimated from the two methods is shown in Fig.\,\ref{fig:mass_wd}. The mean mass of the DA white dwarfs (entire sample) is $0.61\pm0.13\,\Msun$ from La Plata fits ($0.60\pm0.15\,\Msun$ from Montreal) and agrees well with the reported values in literature \citep{Tremblay2019_param, Kilic2020}. Uncertainties in this section correspond to the standard deviation. We notice that the mass distribution of the full sample cannot be fitted by a single Gaussian, hence we performed double Gaussian fits to better illustrate its properties. We find that the distribution exhibits a main peak located at $0.54 \pm 0.05\,\Msun$ (Montreal) and $0.56 \pm 0.03\,\Msun$ (La Plata) with a secondary broad peak at the tail of the distribution at $0.80 \pm 0.08\,\Msun$ (Montreal) and $0.77 \pm 0.08\,\Msun$ (La Plata). Note that the objects with masses $>0.7\,\Msun$ could be over-represented in our sample since they were explicitly targeted in Cycle\,25 (program 15073). Hence, some fraction of this secondary peak at 0.8\,\Msun\ could be due to the sample selection function and may not inherently represent the underlying distribution. 

We also find a small number of low-mass white dwarfs with masses smaller than $0.45\,\Msun$ i.e. 11 (from the La Plata fits) and 14 (from the Montreal fits). Binary interactions are needed to explain their formation as single-star evolutionary models are unable to generate them within the Hubble time. Thus, their masses determined from our fit might not be the true masses if there are two unresolved white dwarfs. We have excluded these low-mass white dwarfs while calculating the double Gaussian fit parameters. 

One of the hypotheses for the secondary peak in the mass distribution is that the massive peak is likely to be formed through the mergers of white dwarfs in binary systems \citep{Liebert2005, Kleinman2013, Rebassa2015,Kilic2018}, however, \cite{Tremblay2016} concluded that there is no direct evidence of the population of double white dwarf mergers in their observed mass distributions. More recently, based on binary population synthesis models \citep{temmink2020}, it was demonstrated by \cite{Kilic2020} that the single white dwarfs formed from mergers cannot entirely explain the peak of intermediate-mass white dwarfs seen in the mass distribution of their 100\,pc sample. An alternative explanation given by \cite{Tremblay2016} and \cite{Kareem2018} is that the secondary peak is produced due to the flattening of initial-final mass relation (IFMR) at initial masses $3.5\leq\rm{M}/\Msun\leq 5.5$ 
with a wide range of them accumulating at white dwarf masses $\sim$0.8\,\Msun. 
Another possible explanation is the delay in cooling due to the release of latent heat from crystallisation that can result in the pile-up of massive white dwarfs \citep{Kilic2020}. However, this is not relevant in our sample because the vast majority are not massive enough (only five stars with $\geq1.0\,\Msun$) to have started core crystallisation given their relatively warm temperatures ($\geq$15\,000\,K). 

\subsection{Mass distribution variation with distance and reddening}
\begin{figure}
\centering
\includegraphics[width=\columnwidth]{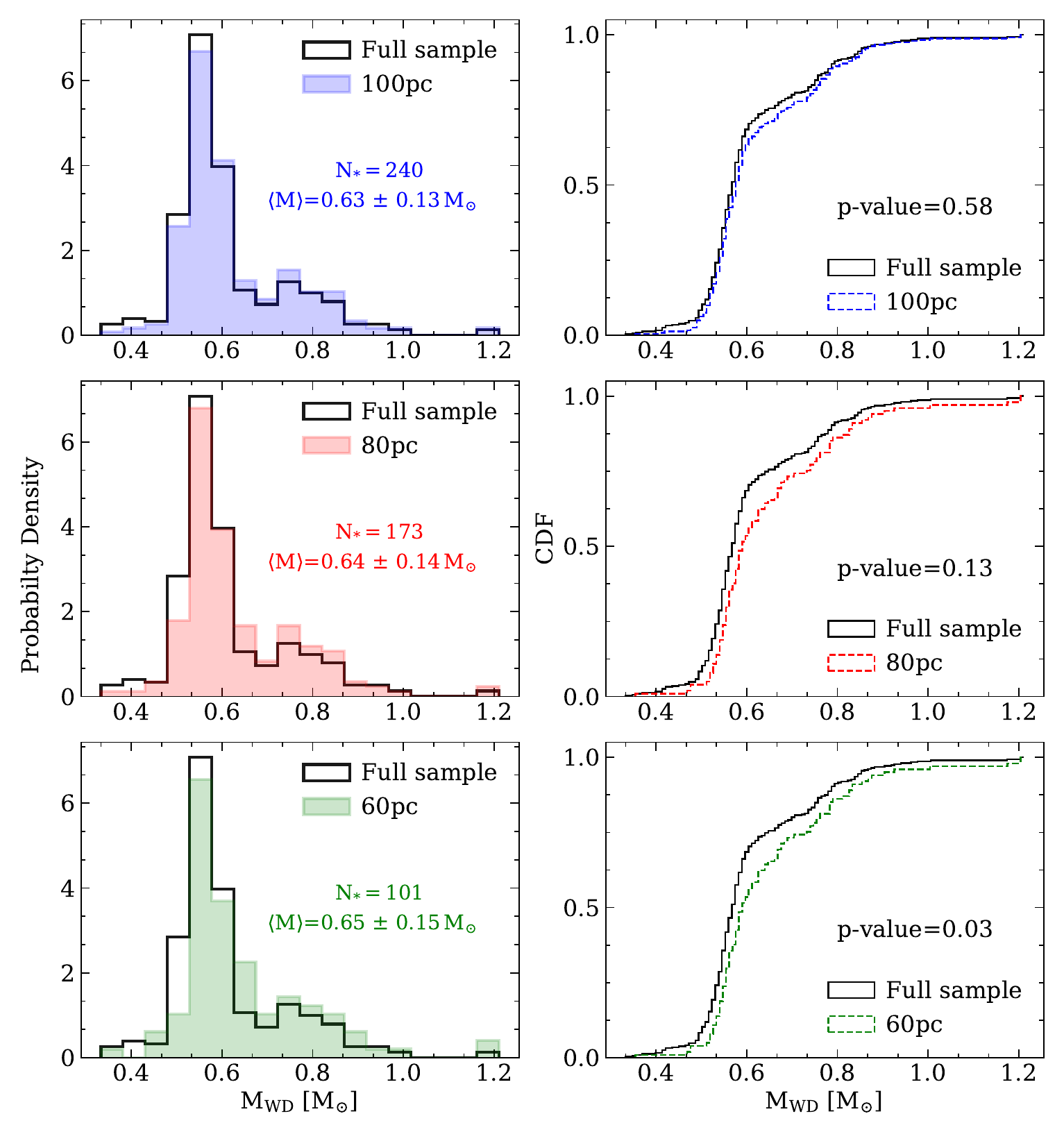}
\caption{Probability density (left) and cumulative distribution functions (CDF, right) of the white dwarf masses based on our fits to the COS spectra using the La Plata M-R relation for the full sample (black) compared with 100\,pc (blue), 80\,pc (red) and 60\,pc (green) samples as shown in upper, middle, and lower panels, respectively. The number of common stars, mean mass (left), and the p-values obtained from the K-S test (right) are marked in the figures. Smaller distance limits result in more complete samples, hence larger mean white dwarf masses.} 
\label{fig:mass_wd_dist}
\end{figure}

To check how the mass distribution varies with the sample selection, we show the probability density\footnote{Defined as the number of stars in each bin divided by the total number of stars and bin width such that the area under the histogram integrates to 1. See the \href{https://matplotlib.org/stable/api/_as_gen/matplotlib.pyplot.hist.html}{matplotlib documentation} for more details.} and cumulative distribution functions of the full COS sample and the sub-samples limited for distances of $<100$, $<80$ and $<60$\,pc in Fig.\,\ref{fig:mass_wd_dist} (La-Plata M-R fits). Given that the sample selection is based on a S/N cut, the figure indicates that as we go out as a function of distance the high mass i.e. low luminosity white dwarfs start dropping out of the sample. Thus, the mean mass of the distribution slightly shifts from a higher value of 0.65\,\Msun\ for 60\,pc to a lower value of 0.61\,\Msun\ for 100\,pc. This is supported by Kolmogorov-Smirnov (KS) test which shows that there is a statistically significant difference in the distribution as the sample size decreases. Specifically, for the 60\,pc sample, the p-value is $\approx0.03$, indicating a significant difference from the full sample distribution, 
while it diminishes with a p-value of 0.13 and 0.58 as we expand to 80 and 100\,pc, respectively. Overall, the shape of the mass distribution remains the same irrespective of volume cuts suggesting that the broad secondary peak is not caused by selection biases. 

\begin{figure}
\centering
\includegraphics[width=\columnwidth]{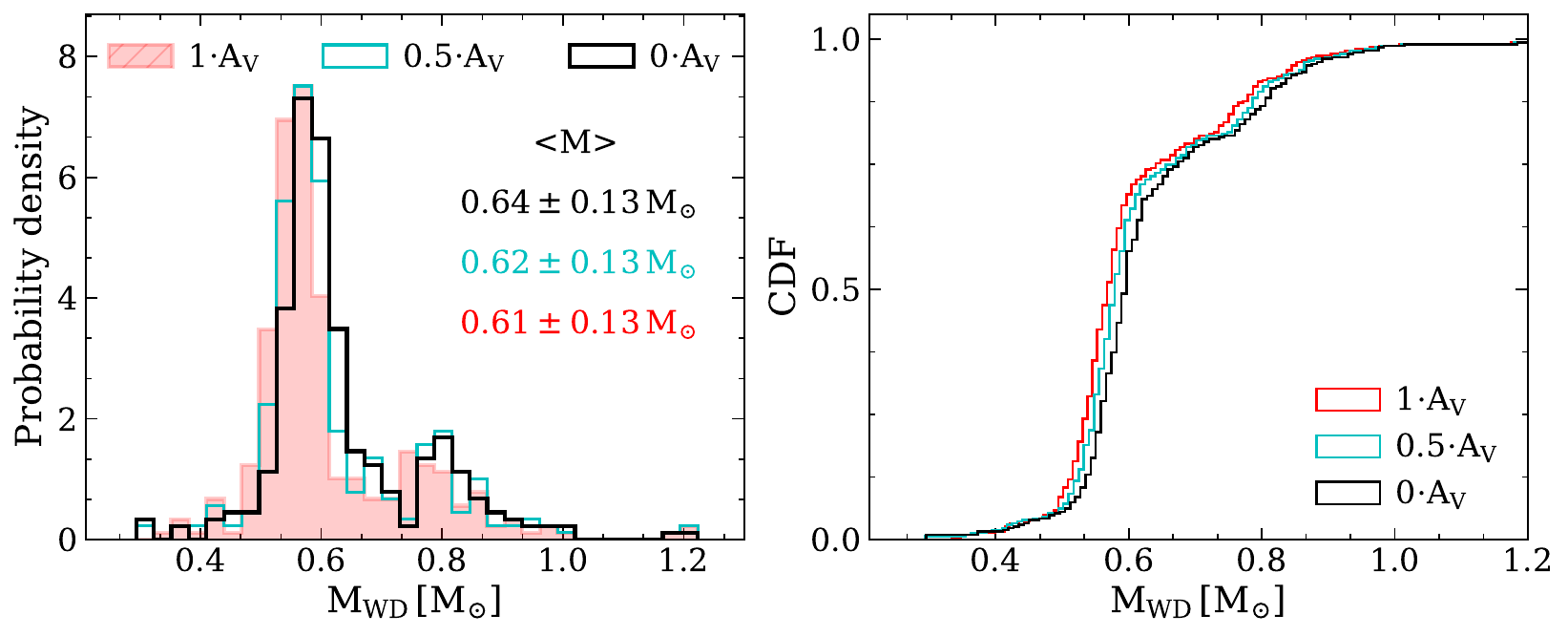}
\caption{Probability density (left) and cumulative distribution functions (CDF, right) of the white dwarf masses based on our fits to the COS data using the La Plata M-R relation, varying the extinction by 0, 0.5 and 1 times the nominal $\rm{A_{V}}$ value as shown in black, cyan, and red, respectively. The mean masses and standard deviation are labelled in the figure.} 
\label{fig:mass_wd_av}
\end{figure}

Since interstellar extinction is more prominent in shorter wavelength regions compared to the optical range (assuming a $\approx 1/\lambda$ dependence), it can significantly alter the shape of the UV flux distribution. To investigate its impact on the mass distribution, we refitted the COS spectra considering two scenarios: assuming no extinction and assuming 0.5 times the A$\rm{_V}$ values in the model spectra. The resulting distributions are shown in Fig.~\ref{fig:mass_wd_av}. We notice that the mean mass shifts from 0.61 to 0.64\,\Msun\ with the masses being systematically higher if we do not account for extinction. 
This suggests that interstellar reddening has a significant consequence in the mass estimates in UV even for the sources lying within 100\,pc and thus can not be ignored while deriving the parameters from UV observations. 

\subsection{Mass distribution comparison with different studies}

\begin{figure}
\centering
\includegraphics[width=\columnwidth]{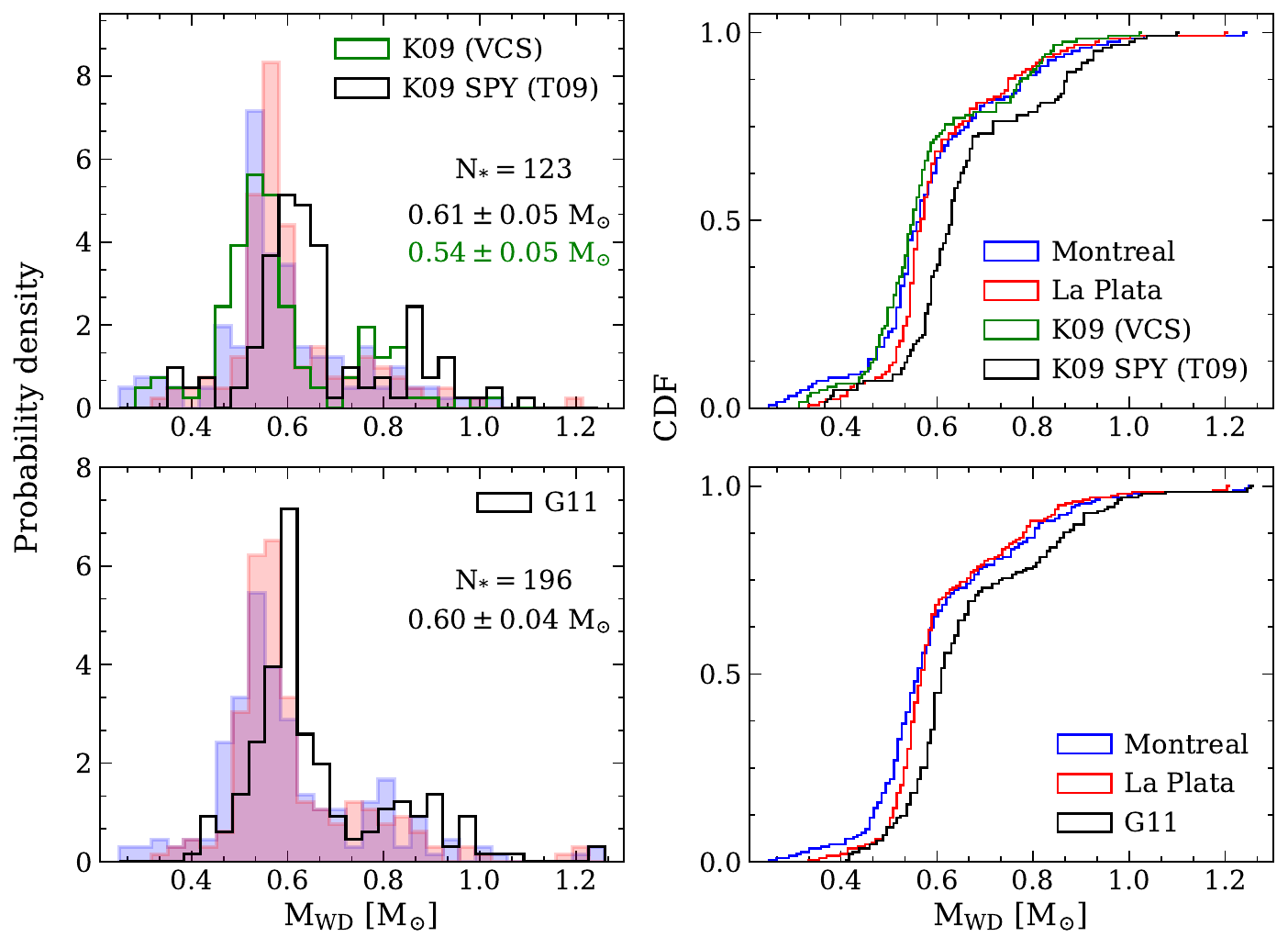}
\caption{Probability density (left) and cumulative distribution functions (CDF, right) of the white dwarf masses based on the fits to the COS spectra using the La Plata (red) and Montreal (blue) M-R for samples common with K09 (top panels, fits using VCS and TB09 Stark profiles shown in green and black, respectively), and G11 (bottom panels, black). The number of common stars and peak masses of the studies obtained from Gaussian fits are reported in the left panels.} 
\label{fig:mass_wd_spec}
\end{figure}

\begin{figure}
\centering
\includegraphics[width=\columnwidth]{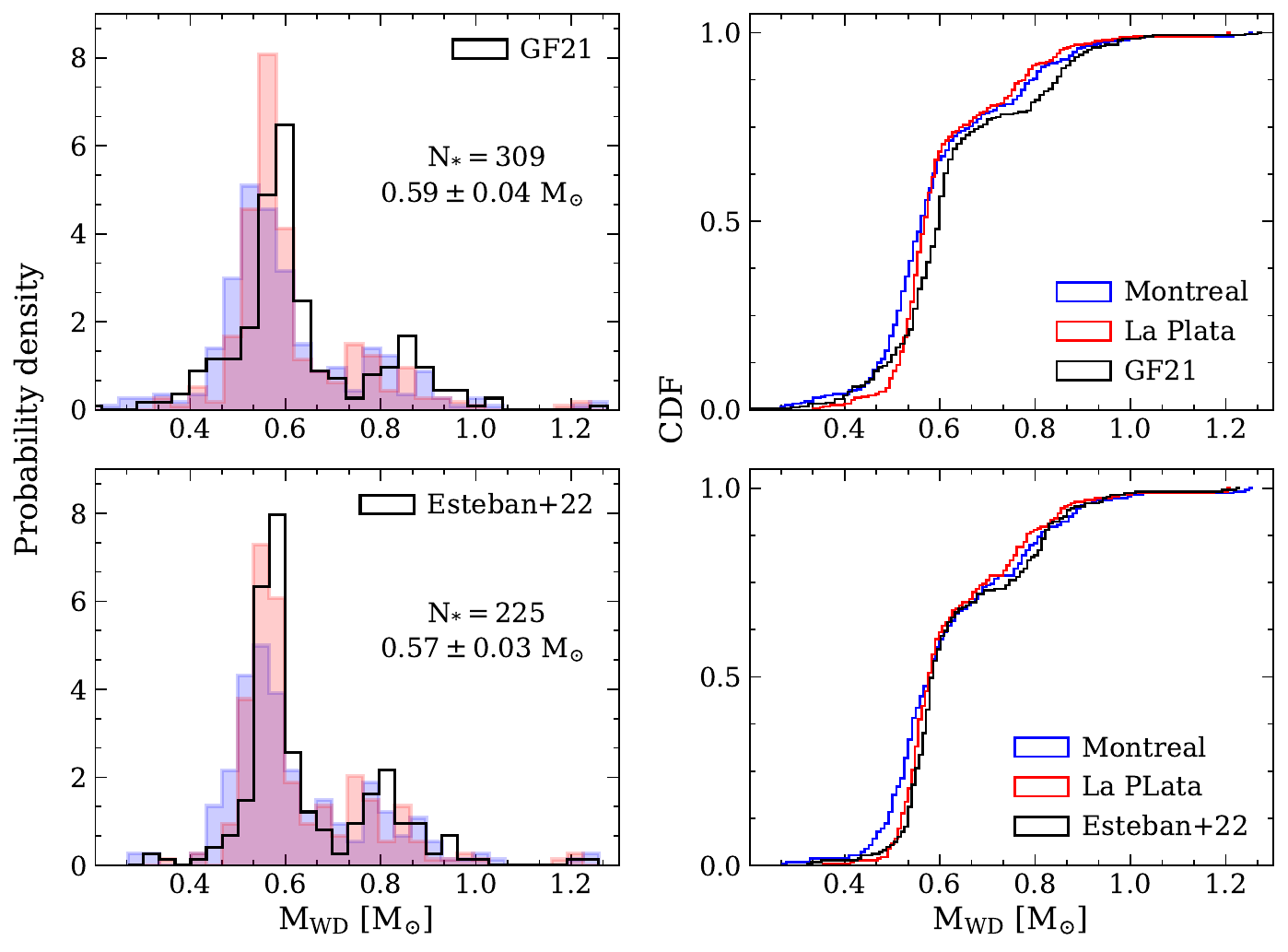}
\caption{Same as in Fig.\,\ref{fig:mass_wd_spec} but for mass comparisons between our COS results with those of the optical photometric studies from GF21 (top panels) and \citet{Esteban2022} (bottom panels).} 
\label{fig:mass_wd_phot}
\end{figure}

In general, the mass distribution studies of white dwarfs differ whether the sample is magnitude-limited, volume-limited, or in-between \citep{Tremblay2016}, and thus, comparing mean or median masses may not give meaningful results. Hence, we selected only the stars in common for comparison with previous literature. Specifically, we compared the COS mass distribution with the spectroscopic studies of K09 and G11 having 123 and 196 common stars, respectively, and photometric studies of GF21 and \cite{Esteban2022} where we found 309 and 225 stars in common with our sample, respectively. Figs.\,\ref{fig:mass_wd_spec} and \ref{fig:mass_wd_phot} depict that the mass distributions of these studies are similar to COS with a secondary peak in the high mass end. However, differences are noticeable in terms of mass shifts with the main peak of the mass distribution being lower in COS ($0.54/0.56\pm0.05/0.03\,\Msun$ for Montreal/La Plata M-R relations) compared to Balmer line fits ($\approx0.60\pm0.05\,\Msun$). In the case of K09 (Fig.\,\ref{fig:mass_wd_spec}, top), the COS masses are higher by $\approx$0.02\,\Msun\ than the masses obtained from the earlier models with VCS Stark profiles. In the case of K09 (with updated TB09 profiles) and G11, the COS masses are lower with a shift of $+$0.05\,\Msun.

Compared to the photometric study of GF21, the main peak lies at $0.59\pm0.05\,\Msun$ as shown in Fig.\,\ref{fig:mass_wd_phot} (left panel), with an overall mass shift of $+0.03\,\Msun$. Considering a more recent study by \cite{Esteban2022} which uses the \textit{Gaia}\,DR3 spectra (integrated to JPAS photometry) and La Plata models, the mass shift is $+0.02\,\Msun$. We found their mass distribution is in good agreement with COS mainly for La Plata fits having a p-value of 0.14 (from KS test), as shown in Fig.\,\ref{fig:mass_wd_phot} (right panel). The cumulative distribution plots also suggest that the UV masses obtained from La Plata fits are in close agreement with \cite{Esteban2022}, for the masses $\leq0.7\,\Msun$. While for masses higher than this, the Montreal fits agree better.

\section{Outliers}\label{sec:outliers}
We flagged the outliers based on three different methods: (a) poor fit to the COS spectra, (b) large disagreements between the COS \Teff\ and \logg\ with previous spectroscopic and photometric studies, and (c) known binaries including white dwarf-main sequence companions and double degenerates. The summary of the outliers is provided in Table\,\ref{tab:outliers_summ}. 

\begin{table*}
\centering
\caption{Summary of outliers where mass estimates are based on La~Plata M-R relations (Montreal in brackets). The $\chi_r^2$ is based on the model fit to the COS spectra. The columns phot-\Teff, phot-\logg, correspond to the outliers based on the comparisons of COS parameters with photometric studies \citep{nicola2021} while column spec-\logg\ represents the outliers with respect to spectroscopic studies (G11). We refer to Section\,\ref{sec:outliers} for more details on their selection.}
\begin{tabular}{cccccccc} \hline
Object	&	Mass (\Msun) & $\chi_r^2$ &	phot-\Teff	&	phot-\logg	&	spec-\logg	&	RUWE	&	Comments	\\\hline
\multicolumn{8}{c} {from comparative analysis}\\\hline
HS\,0200+2449	&	0.75	(	0.78	)	&	1.14	&	\Y	&	\Y	&	\N	&	1.00	&	$\dagger$	\\
HS\,1334+0701	&	\textit{0.43}	(	\textit{0.34}	)	&	0.74	&	\Y	&	\Y	&	\N	&	1.02	&	DDs$^1$	\\
HS\,2220+2146A	&	\textit{0.31}	(	\textit{0.28}	)	&	0.31	&	\N	&	\N	&	\Y	&	1.03	&		\\
PG\,1220+234	&	0.70	(	0.72	)	&	1.04	&	\N	&	\N	&	\Y	&	1.07	&		\\
WD\,0028$-$474	&	0.51	(	0.47	)	&	0.80	&	\Y	&	\Y	&	\N	&	1.03	&	DDd$^1$	\\
WD\,0136+768	&	0.52	(	0.49	)	&	0.73	&	\Y	&	\Y	&	\N	&	1.14	&		\\
WD\,0231$-$054	&	0.67	(	0.68	)	&	0.60	&	\N	&	\N	&	\Y	&	0.94	&		\\
WD\,0321$-$026	&	0.52	(	0.47	)	&	0.78	&	\N	&	\N	&	\Y	&	0.99	&	$\dagger$, magnetic (<1\,MG)$^4$	\\
WD\,0437+152	&	\textit{0.33}	(	\textit{0.25}	)	&	0.85	&	\N	&	\Y	&	\N	&	1.08	&	$\dagger$	\\
WD\,0732$-$427	&	1.21	(	1.25	)	&	0.81	&	\Y	&	\Y	&	\Y	&	1.06	&	$\dagger$	\\
WD\,1115+166	&	0.70	(	0.72	)	&	1.33	&	\Y	&	\Y	&	\N	&	1.04	&	*, DDd$^1$	\\
WD\,1230$-$308	&	0.51	(	0.46	)	&	1.04	&	\N	&	\N	&	\Y	&	1.03	&	\\
WD\,1349+144	&	\textit{0.36}	(	\textit{0.27}	)	&	0.96	&	\N	&	\N	&	\Y	&	0.97	&	*, DDd$^1$	\\

WD\,1713+332	&	\textit{0.42}	(	\textit{0.33}	)	&	0.93	&	\Y	&	\N	&	\N	&	1.15	&		\\
WD\,1739+804	&	0.53	(	0.50	)	&	1.06	&	\Y	&	\Y	&	\N	&	0.99	&		\\
WD\,1943+163	&	0.56	(	0.55	)	&	1.04	&	\Y	&	\N	&	\N	&	1.06	&		\\
WD\,2009+622	&	0.51	(	0.48	)	&	1.26	&	\Y	&	\Y	&	\N	&	0.93	&		\\
WD\,2200$-$136	&	0.50	(	0.46	)	&	0.97	&	\Y	&	\Y	&	\N	&	1.11	&	DDd$^1$	\\

WD\,2359$-$324	&	0.52	(	0.48	)	&	0.90	&	\Y	&	\Y	&	\N	&	1.02	&		\\
WD\,J015630.05+295532.28 & 0.86 ( 0.89 ) & 0.79 & \N & \Y & \N & 1.07 & \\
WD\,J074152.84$-$570844.74	&	0.51	(	0.47	)	&	1.19	&	\N	&	\N	&	\Y	&	1.12	&	*, Binary$^3$ 	\\
WD\,J155501.99+351328.65	&	0.55	(	0.54	)	&	1.42	&	\Y	&	\Y	&	\N	&	1.11	&		\\

WD\,J175151.11$-$202308.72	&	0.68	(	0.69	)	&	1.03	&	\Y	&	\N	&	\N	&	1.07	&		\\
WD\,J180240.42$-$243603.86	&	0.56	(	0.55	)	&	1.02	&	\Y	&	\N	&	\N	&	1.06	&		\\
WD\,J181058.67+311940.94	&	\textit{0.35}	(	\textit{0.27}	)	&	1.21	&	\Y	&	\Y	&	\N	&	1.13	&	*	\\
WD\,J182315.21+170639.42	&	0.53	(	0.50	)	&	1.05	&	\N	&	\Y	&	\N	&	1.05	&		\\
WD\,J202359.51$-$422425.85	&	0.76	(	0.78	)	&	1.63	&	\Y	&	\Y	&	\N	&	1.04	&	$\dagger$\\	
APASS\,J195622.94+641358.0	&	0.52	(	0.50	)	&	0.81	&	\Y	&	\Y	&	\N	&	0.95	&		\\\hline
\multicolumn{8}{c} {high RUWE ($>1.4$)}\\\hline
HE\,0131+0149	&	0.55	(	0.57	)	&	0.74	&	\Y	&	\N	&	\N	&	4.16 &	$\dagger$, DDs$^1$	\\
HE\,2218$-$2706	&	0.54	(	0.52	)	&	0.71	&	\Y	&	\N	&	\N	&	6.00 &		\\
HE\,2231$-$2647	&	0.60	(	0.60	)	&	0.93	&	\Y	&	\Y	&	\N	&	2.61 &	\\ 
PG\,2345+305	&	0.54	(	0.52	)	&	1.15	&	\Y	&	\N	&	\N	&	2.21 &		\\
WD\,0216+143	&	0.60	(	0.60	)	&	1.27	&	\Y	&	\Y	&	\N	&	2.76 &	DDs$^1$	\\
WD\,1129+155	&	0.58	(	0.59	)	&	0.98	&	\Y	&	\Y	&	\N	&	6.49 &	\\ 
WD\,1531$-$022	&	0.48	(	0.42	)	&	1.10	&	\N	&	\Y	&	\Y	&	2.88 &	*, possibly composite$^8$, DD?	\\
WD\,2328+107	&	0.59	(	0.59	)	&	0.94	&	\Y	&	\N	&	\N	&	3.38 &	circumstellar disc$^7$	\\
WD\,J141039.06$-$474439.48	&	0.61	(	0.62	)	&	1.23	&	\Y	&	\N	&	\N	&	5.01 &	Binary (RV variable)$^6$\\ 
WD\,J170909.53+473134.68	&	0.58	(	0.58	)	&	1.36	&	\Y &	\Y	&	\N &	4.29 &	$\dagger$ \\ 
WD\,J055905.17+022802.50 & 0.50 ( 0.46 ) & 1.21 & \N & \N & \N & 1.42 & $\dagger$\\ 
WD\,0920+363 & \textit{0.44} ( \textit{0.36} ) & 0.95 & \N & \N & \N & 1.74 & low mass\\\hline 
\multicolumn{8}{c} {known binaries or other systems (not in the above-mentioned selection criteria)}\\\hline
WD\,0128$-$387	&	0.63	(	0.63	)	&	0.66	&	\N	&	\N	&	\N	&	1.03	&	DDd$^1$, smeared H$_2^+$ feature	\\
WD\,0341+021	&	\textit{0.30}	(	\textit{0.37}	)	&	0.99	&	\N	&	\N	&	\N	&	1.07	&	$\dagger$, DDs$^1$	\\
WD\,0843+516	&	0.58	(	0.57	)	&	1.40	&	\N	&	\N	&	\N	&	0.94	&	circumstellar disc$^2$	\\
WD\,1015+161	&	0.59	(	0.58	)	&	0.88	&	\N	&	\N	&	\N	&	1.11	&	circumstellar disc$^2$	\\
WD\,1229$-$013	&	\textit{0.42}	(	\textit{0.34}	)	&	0.98	&	\N	&	\N	&	\N	&	1.19	&	low mass	\\
WD\,1249+160	&	\textit{0.41}	(	\textit{0.32}	)	&	1.32	&	\N	&	\N	&	\N	&	1.09	&	low mass	\\
WD\,1555$-$089	&	0.56	(	0.54	)	&	0.74	&	\N	&	\N	&	\N	&	1.00	&	CPM binary$^5$	\\
WD\,1929+011	&	0.71	(	0.72	)	&	5.36	&	\N	&	\N	&	\N	&	1.14	&	circumstellar disc$^2$	\\
WD\,2032+188	&	\textit{0.41}	(	\textit{0.32}	)	&	0.84	&	\N	&	\N	&	\N	&	1.08	&	DDs$^1$	\\
HE\,2345$-$4810	&	\textit{0.43}	(	\textit{0.35}	)	&	1.25	&	\N	&	\N	&	\N	&	1.01	&	DDs$^1$	\\
WD\,J055635.50$-$561006.57	&	0.70	(	0.72	)	&	1.08	&	\N	&	\N	&	\N	&	0.99	&	*	\\
WD\,J150156.33+302258.23	&	0.55	(	0.52	)	&	1.41	&	\N	&	\N	&	\N	&	1.15	&	Binary$^3$ (DA+K/M)	\\\hline
\label{tab:outliers_summ}
\end{tabular}
\begin{tablenotes}\footnotesize
    \small
    \item Notes: \Y\ denotes that the target is an outlier in the respective category, whereas, \N\ denotes otherwise. Targets having masses $\leq0.45\,\Msun$ are shown in italics. DD: Double Degenerate where DDs and DDd denote a single-lined and double-lined spectroscopic binary, respectively, CPM: Common proper motion binary, $*$: Ly$\alpha$ core not well fit, $\dagger$: Blue wing of Ly$\alpha$ ($<1200$\AA) does not fit well\\
    References:
    $^1$\cite{Koester2009}, $^2$\cite{boris2012}, 
    $^3$\cite{mccook1999}, 
    $^4$\cite{ferrario2015},
    $^5$\cite{1991ApJ...375..674W},
    $^6$\cite{2000MNRAS.319..305M},\\
    $^7$\cite{2015MNRAS.449..574R},
    $^8$\cite{Napiwotzki2020}
\end{tablenotes}
\end{table*}

\subsection{Poor fits to COS spectra}
\label{sec:poor_fits}
We find that $\approx15$\,per cent of the stars in our sample have bad $\chi^{2}_\mathrm{r}<0.7$ or $>1.2$. As the $\chi^{2}$ is weighted by the errors on observed fluxes, the reason for a very large or small $\chi^{2}_\mathrm{r}$ could be either due to (1) the underestimation or overestimation of the errors, or due to (2) the real deviation from the model fit due to an intrinsic reason. Hence, we closely examined their spectra and model fit. 

\begin{figure*}
\centering
\includegraphics[width=\textwidth]{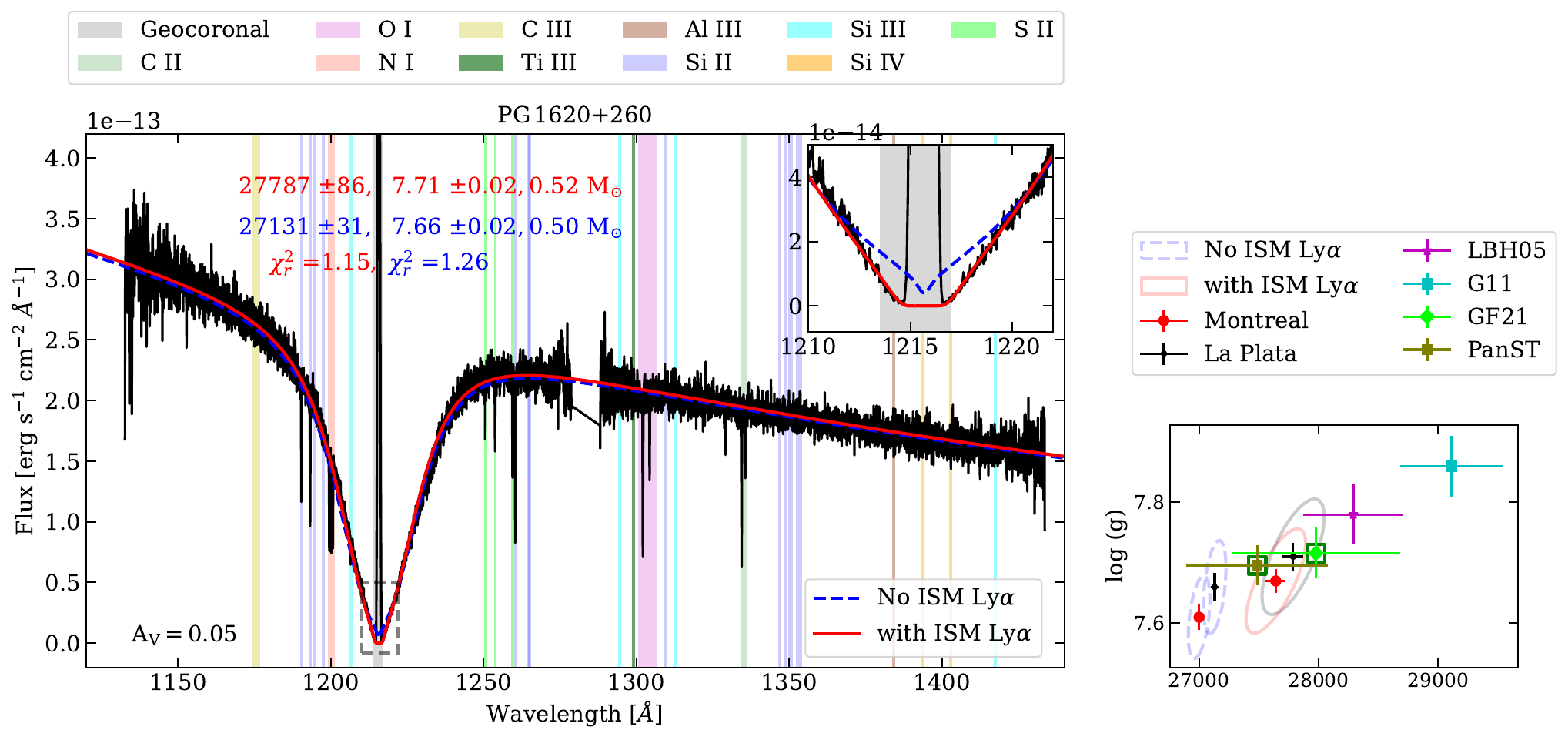}
\caption{Left panel: The model fit to the COS spectrum of PG\,1620+260 (with $A_V=0.05$) illustrates the effect of including the contribution of interstellar Ly$\alpha$ absorption ($\rm{N (H\,I)}=7.95\times10^{19}\,\rm{cm}^{-2}$). Shown as red solid line is the fit including interstellar Ly$\alpha$, and as blue dashed line the fit without the interstellar component. The zoomed inset (indicated by the grey box) in the top right corner shows the model fit to the core of Ly$\alpha$. The best fit values, \Teff, \logg, mass, $\rm{\chi^{2}_{r}}$ from both the cases (red: with ISM Ly$\alpha$, blue: without ISM Ly$\alpha$) are labelled in the figure. Right panel: The corresponding 95\,per cent confidence ellipses (red and grey for two M-R relations) show that the \Teff\ and \logg\ are slightly higher when ISM Ly$\alpha$ is considered in the fit than when it is not (blue dashed ellipses). The literature values from \citet{Liebert2005} (LBH05), \citet{Gianninas2011} (G11), \citet{nicola2021} (GF21), and from Pan-STARRS data (PanST) derived in this work are shown in the figure.  Refer to Fig.\,\ref{fig:spec_fit_wd_good} for a more detailed description of symbols.}
\label{fig:ism_lya}
\end{figure*}

Neutral hydrogen along the line-of-sight will cause interstellar Ly$\alpha$ absorption in the observed white dwarf spectra. The neutral hydrogen column density is well correlated with reddening, $E(B-V)$, \citep{1994ApJ...427..274D} and for the range of reddening of the COS sample, this mainly affects the core of the Ly$\alpha$. Reddening is generally larger for more distant stars, which in our flux-limited sample will affect stars hotter than 24\,000\,K. Inspecting the fits of the hottest stars in our sample, we note that 23 of them have large $\chi^{2}_\mathrm{r}\geq1.2$. Among these, ten stars have a broadened Ly$\alpha$ core which does not fit well by the model. We re-performed the fit adding the contribution of ISM Ly$\alpha$ absorption in the model using the relation $\mathrm{N(H\,I)}=4.93\times10^{21}\times E(B-V)\,\mathrm{[cm^{-2}]}$ \citep{1994ApJ...427..274D}. We find that the fit improved in terms of  $\chi^{2}_\mathrm{r}$ as shown in Fig.\,\ref{fig:ism_lya}. In addition, the \Teff\ and masses are found to be on average higher by $\simeq500$\,K and 0.02\,\Msun\ respectively, and in better agreement with the literature studies.

\begin{figure*}
\centering
\includegraphics[width=0.9\textwidth]{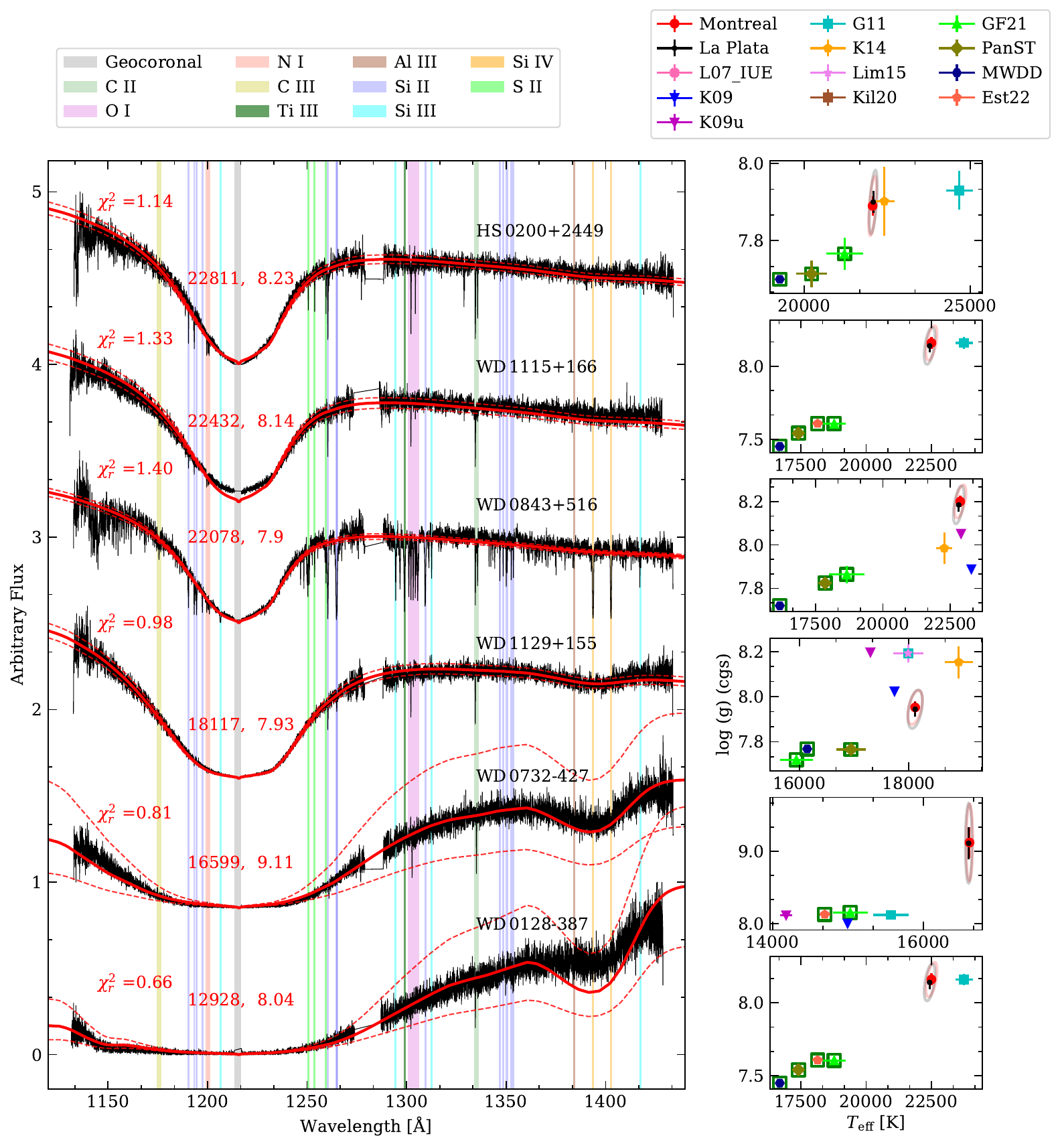}
\caption{Similar to Fig.\,\ref{fig:spec_fit_wd_good} but for white dwarfs with poor model fits ($\chi^{2}_\mathrm{r}>1.2$ for WD\,1115+166, WD\,0843+516) and/or large spread in published \Teff\ and \logg\ (which is the case for all the six stars shown here). It is apparent that not all stars where we find a large disagreement with the published atmospheric parameters also have poor COS fits (e.g. WD\,1129+155). HS\,0200+2449 and WD\,0732$-$427 have poor model fits in the blue end of the spectrum ($\lambda\la1170$\,\AA), while the COS spectrum of WD\,0128$-$387 is poorly fit in the H$^+_2$ region (1380$-$1410\,\AA). The physical reasons for the bad fits and/or the spread in atmospheric parameters are that these systems are either confirmed (WD\,0128$-$387, WD\,1115+166) or suspected (WD\,0732$-$427, WD\,1129+155) double-degenerates, or have large amounts of metals in their atmospheres (WD\,0843+516). The case of HS\,0200+2449 is not clear.}
\label{fig:spec_fit_wd_bad}
\end{figure*}

Through visual inspection, we noticed that the core of the Ly$\alpha$ region is not fit well by the DA models in six stars that cannot be explained by the ISM Ly$\alpha$ absorption (e.g. WD\,1115+166, Fig.\,\ref{fig:spec_fit_wd_bad}).  Among these systems are two known double degenerates (WD\,0341$+$021, WD\,1115$+$166), WD\,J074152.84$-$570844.74 is a known binary \citep{mccook1999},  WD\,1531$-$022 has been classified as a possible composite system in the SPY survey \citep{Napiwotzki2020}. The other two systems with poor Ly$\alpha$ fits are the WD\,J055635.50$-$561006.57, WD\,J181058.67+311940.94, making them strong double-degenerate candidates. Both stars currently have only COS spectroscopy, and optical time-series spectroscopy will be required to probe for radial velocity variations. In the case of WD\,0128$-$387 the H$^{+}_{2}$ satellite feature is smeared out in the COS spectrum as clearly visible in Fig.\,\ref{fig:spec_fit_wd_bad}, thus the DA models do not fit well in that region. This spectroscopic morphology can be explained by the presence of a white dwarf companion that is not of DA type, which supports the classification of this system as a DA+DB by K09. 

There are eight cases where the model atmospheres do not fit well the blue end ($\lambda\la1200$\,\AA) of the COS spectrum, i.e. the blue wing of Ly$\alpha$, e.g. HS\,0200$+$2449, which shows a large scatter in the published atmospheric parameters (Fig.\,\ref{fig:spec_fit_wd_bad}). Similarly, the fit is bad for WD\,0732$-$427, especially in the Ly$\alpha$ core and its blue wing (see Fig.\,\ref{fig:spec_fit_wd_bad}). We determine a very high mass for this star, $\simeq1.2$\,\Msun, making it a clear outlier with respect to the published spectroscopic and photometric studies, which all report a lower mass ($\simeq0.7\,\Msun$). We conclude that WD\,0732$-$427 is most likely an unresolved double-degenerate, in which the hotter and more massive component dominates the UV flux.

Finally, in two of the sources, the UV continuum is affected by the presence of numerous strong metal absorption lines, thus resulting in a poor fit and hence large $\chi^{2}_\mathrm{r}$: WD\,0843+516 (Fig.\,\ref{fig:spec_fit_wd_bad} with $\chi^{2}_\mathrm{r}=1.4$) and WD\,1929$+$011 ($\chi^{2}_\mathrm{r}=5.4$). Both stars have detected circumstellar discs from which material accretes into the white dwarf atmospheres and are classified as DAZ \citep{boris2012}. The fits of these stars can be improved by adopting the same methodology but adding a metal absorption line mask or fitting the continuum and metal lines together.

\subsection{Photometric and spectroscopic outliers}

\begin{figure}
\centering
\includegraphics[width=\columnwidth]{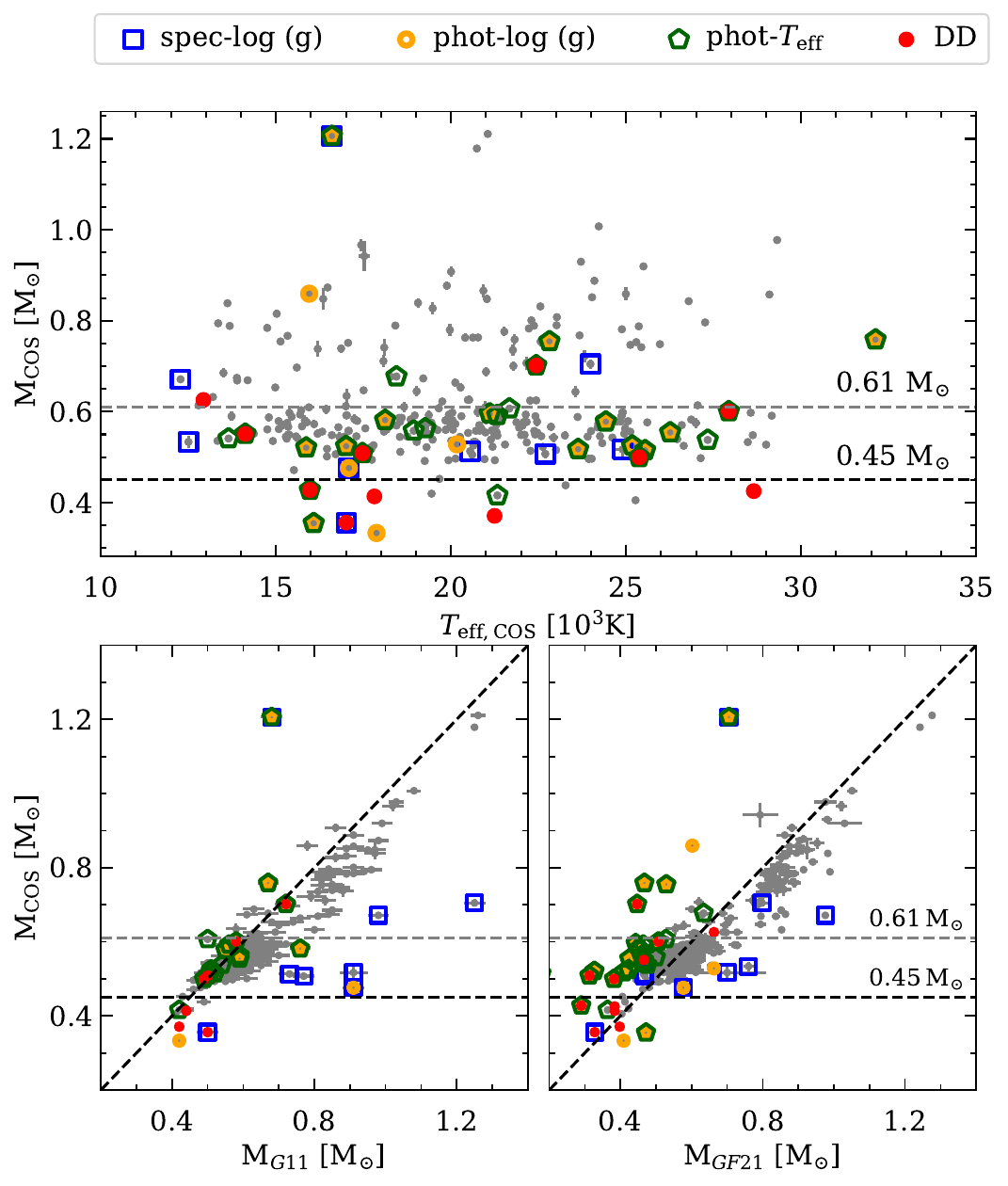}
\caption{Top panel: mass vs \Teff\ for the COS sample from the fits using the La Plata M-R. Bottom panel: Masses from the COS analysis versus those from G11 (left) and GF21 (right). The outliers identified from spectroscopic and photometric comparisons of \logg\ are marked in blue squares and orange circles, and from photometric comparisons of \Teff\ in green pentagons, respectively.}
\label{fig:outl_plot}
\end{figure}

We define outliers identified from comparisons with previous studies as systems having an absolute \Teff\ difference of $\geq5$\,per cent between COS fits and GF21 (27 \textit{photometric $\mathit{T_{eff}}$ outliers}), a difference of $\pm0.15$\,dex between COS and GF21 (22 \textit{photometric $\mathit{log g}$ outliers}), and, a difference of $\pm0.3$\,dex between COS and Balmer line fits (9 \textit{spectroscopic $\mathit{log g}$ outliers}), considering $2\sigma$ errors. 18 of the 22 photometric \logg\ outliers are also photometric \Teff\ outliers. One system, the massive double-degenerate candidate WD\,0732$-$427 discussed in Sect.\,\ref{sec:poor_fits} falls in all three categories. The 37 outliers are shown in the \Teff\ vs mass plane in the top panel of Fig.\,\ref{fig:outl_plot}, and a comparison of their masses measured from the COS spectra with those determined from optical spectroscopy (G11) and photometry (GF21) are shown in the bottom left and right panels of Fig.\,\ref{fig:outl_plot}, respectively. We note that the majority of the outliers ($\approx$80\,per cent) have UV masses less than the mean mass ($0.6\,\Msun$) of the COS sample and are randomly distributed at all effective temperatures. Among them, five have masses smaller than $0.45\,\Msun$, suggesting their formation through a binary channel. 

Among the spectroscopic outliers in \logg\ (excluding WD\,0732$-$427, see above), WD\,1531$-$022 and WD\,0740$-$570 are known composite systems while the rest (HS\,2220+2146A, WD\,0321$-$026, PG\,1220+234, WD\,1230-308, WD\,0231$-$054, WD\,1349+144) have larger masses based on the Balmer line fits (G11 and K09) when compared to the results obtained from the COS analysis, as shown in the lower left panel of Fig.\,\ref{fig:outl_plot}. This implies that these six systems could be unresolved DA+DA binaries of similar masses. We note that HS\,2220+2146A has a wide white dwarf common proper motion companion, HS\,2220+2146B, with a separation of 6.6\,arcsec, and a projected separation of $\simeq470$\,au. If HS\,2220+2146A is indeed a close double-degenerate, it would make this system a hierarchical triple, similar to WD\,1704+481 \citep{2000MNRAS.314..334M}. In the case of photometric outliers (both \Teff\ and \logg), 90\,per cent have larger masses measured from the UV compared to masses determined from optical photometry (Fig.\,\ref{fig:outl_plot}, lower right panel). In addition, they have higher \Teff\ than the photometric estimates suggesting that some of them could be unresolved binary candidates. 

\begin{figure}
\centering
\includegraphics[width=\columnwidth]{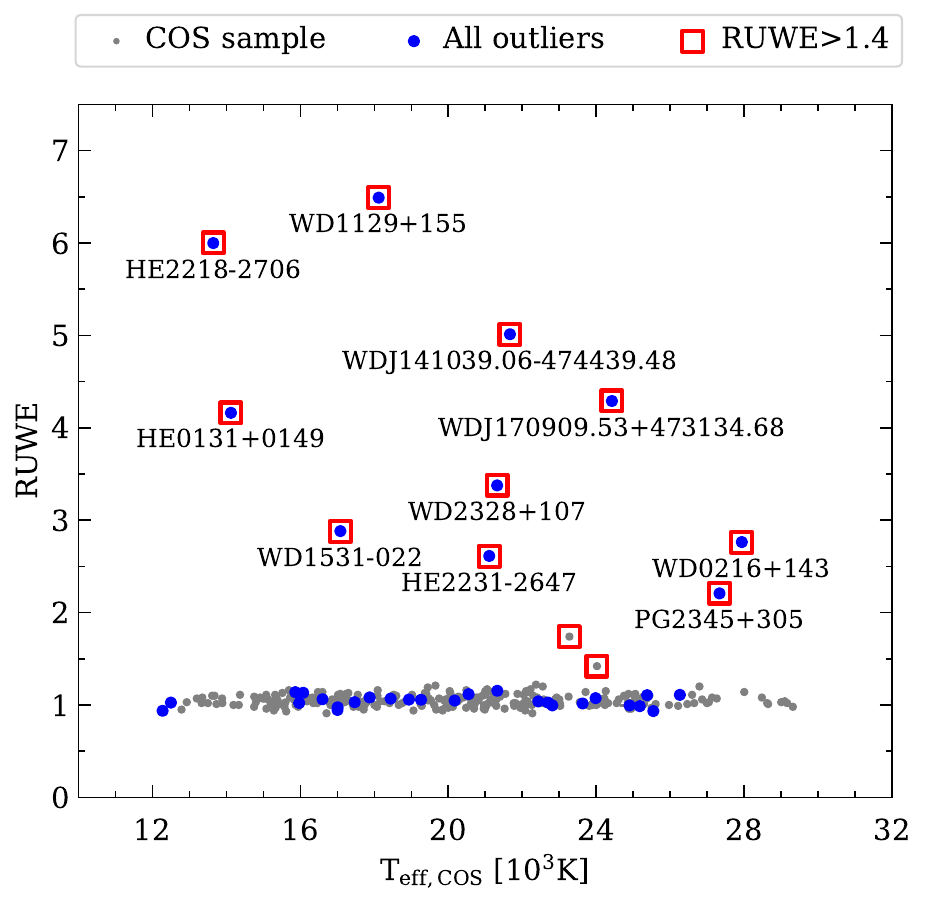}
\caption{\textit{Gaia} RUWE as a function of \Teff\ for the COS sample (grey dots),  where the 37 outliers in our sample (Table\,\ref{tab:outliers_summ}) are marked as blue dots and sources with $\mathrm{RUWE}>1.4$ are highlighted by red squares. The names of the outliers having $\mathrm{RUWE}>2$ are reported in the figure. Refer to Sect.\,\ref{sec:outliers} for more details on the selection of outliers.}
\label{fig:outl_plot_ruwe}
\end{figure}

To investigate further the nature of the selected outliers, we checked the Renormalised Unit Weighted Error (RUWE) parameter from \textit{Gaia}\,DR3 \citep{2021A&A...649A...2L} which is highly sensitive to unresolved binaries. Figure\,\ref{fig:outl_plot_ruwe} shows the RUWE as a function of \Teff\ from La Plata fits for the entire COS sample. According to \cite{LL:LL-124}, well-behaved single sources are expected to have RUWE close to unity as noted for the majority of stars in the sample, whereas the outliers with $\mathrm{RUWE}>1.4$ have poor astrometric fits, hence are probable astrometric binaries. Twelve systems have $\mathrm{RUWE}>1.4$, including ten which are outliers in one or more of the metrics we defined above (Table\,\ref{tab:outliers_summ}). Among these, WD\,1129+155 has the highest RUWE of 6.5 and shows a large spread in the published \Teff\ and \logg\ values (see Fig.\,\ref{fig:spec_fit_wd_bad}) and WD\,0216+143 and HE\,0131+0149 are already known double-degenerates (K09). We conclude that the systems with high RUWE values are likely to be unresolved binaries.

\subsection{Known binaries}
Four of our COS targets are not included in the white dwarf catalogue of GF21  (WD\,0933+025, WD\,0022$-$745, HE\,1117$-$0222, and WD\,1049+103) and are part of wide binaries. Inspecting the Gaia DR3 archive, we found that parallax is available for WD\,0933+025 ($7.465\pm0.096$\,mas) which has an M-dwarf companion with a projected separation of $\simeq1$\,arcsec. Likewise, as WD\,0022$-$745 is a common proper motion pair with an F-type main-sequence companion \citep{1997MNRAS.287..381B}, \textit{Gaia} has a good parallax measurement ($7.676\pm0.013$\,mas) for the companion star, thus, we have used this information in our fitting. Also, we have used the extinction values ($A_{V}=0.05$) derived from 3D STILISM/EXPLORE \citep{lallement2019}. In the case of HE\,1117$-$0222, \textit{Gaia} resolves two stars with very similar colours. Unfortunately, it does not provide parallax, which is why we have excluded it from our sample. Similarly, WD\,1049+103 is resolved by \textit{HST} (separation 0.26\,arcsec) but not by \textit{Gaia}, hence parallax is not available. 

We found 11 known double-degenerate (DD) systems from the SPY survey (K09) in our sample as shown in Fig.\,\ref{fig:outl_plot}, among which five are double-lined spectroscopic systems (DDd; K09), and six are single-lined systems (DDs; K09). Based on our comparative study, six of these systems are photometric outliers having higher UV spectroscopic \Teff\ ($\geq10$\,per cent) and \logg\ ($\geq0.25$\,dex) when compared with the optical photometric estimates. The COS parameters of the other five systems agree with photometric values which indicates that the binaries where both components have similar atmospheric parameters might go undetected from the comparative analysis. Nevertheless, the \logg\ comparisons of four of these systems denote that they have low masses ($<0.45\,\Msun$) as inferred from both the COS and optical studies which suggest that these stars are of binary origin.

\section{Discussion}\label{sec:discus}
Our comparative analysis revealed several discrepancies between the COS results and previous studies, most of which were based on optical data. The \Teff\ obtained from COS fits are found to be consistently lower by on average 3\,per cent and 1.5\,per cent than those from spectroscopic and photometric studies, respectively. Likewise, the COS masses are systematically lower by $\approx$0.052\,\Msun\ ($\pm0.005$)\footnote{Note that these are typical standard errors on the median mass offsets calculated by excluding the outliers} than the masses derived from Balmer line fits and about $\approx$0.024\,\Msun\ ($\pm0.003$) lower than the optical photometric masses. Since \Teff\ and \logg\ are correlated via the M-R relation, parallax, and absolute magnitude, lower masses in COS suggest that we obtain larger radii, which would compensate for the lower \Teff\ we find from the COS analysis compared to the \Teff\ from other studies. To illustrate the correlation between \Teff\ and $M_\mathrm{wd}$, an offset of $-1.5$\,per cent in the COS \Teff\ with respect to the photometric \Teff\ from GF21 would imply an offset of $\approx$0.06\,dex in \logg\ (for constant $L=4\pi R_\mathrm{wd}^{2}\sigma {\Teff}^{4}$). This compares to the 0.07\,dex offset in \logg\ we found between the COS results and those of GF21, suggesting that the mass discrepancy is primarily due to the COS \Teff\ being lower than those from previous studies~--~and not from any issue with the absolute flux calibration of the COS spectroscopy. 

Similarly, a $-5$\,per cent offset between the COS \Teff\ and those based on the Balmer line fits of  G11 implies a 0.2\,dex offset in \logg, compared to the 0.1\,dex offset found between the \logg\ from our  COS analysis and those from G11. This suggests that the offset in $\log g$ has nearly equal contributions from the COS \Teff\ being lower and from an intrinsic difference between spectroscopic and photometric mass determinations. To better understand the possible cause(s) for the lower \Teff\ and masses found from the COS data, we performed the following tests:
\begin{enumerate}

\item \textit{Mass-radius relations}: Since the Montreal and La Plata models assume different core compositions and masses of the H envelopes, they result in significantly different stellar masses, especially for stars with $<0.5\,\Msun$. For the COS sample, the masses are comparatively higher in La Plata fits than Montreal fits in this range, and thus are in better agreement with Balmer line fits, while the opposite is seen for higher mass stars with mass $>0.7\,\Msun$. However, the differences in the COS \Teff\ using the two mass-radius relations are negligible, and cannot fully explain the observed systematic offsets in the fits when compared with previous studies.\\

\item \textit{Model spectra}: Earlier works relied on model spectra \citep{Liebert2005, Lajoie2007, Koester2009} using the unified theory of Stark broadening from VCS \citep{vidal1973}, whereas the later works (G11, K14, GF21) as well as our analysis made use of the TB09 Stark profiles, although this makes little difference for photometric and COS UV fits. Using the updated profiles, we noted the differences in COS and K09 \Teff\ to be reduced by two per cent compared with K09 (VCS profiles). Even the Stark broadening profiles of TB09 remain uncertain \citep{Cho_2022}, and this could possibly explain the systematically higher masses and temperatures found from Balmer line studies \citep{Tremblay2019_param,genest2019}. In the case of K14 UV study, as the same microphysics is used in the models, the \Teff\ offset with their work might be arising due to fitting methods, reddening correction, or changes in the COS data calibration. \\
    
\item \textit{Interstellar reddening}: Using \textit{IUE} data, \cite{Lajoie2007} found that reddening plays an important role in the observed \Teff\ differences with optical studies. Taking reddening into account, we note that the \Teff\ values are lower by $\approx$0.7\,per cent and the mean masses by 0.03\,\Msun\ compared to the values when reddening is neglected. This is an extreme case and the neglect of UV extinction is unlikely to be realistic, as illustrated by the numerous interstellar absorption lines seen in white dwarf COS spectra. Given the similar fitting techniques and input model physics in COS and photometric studies, this nevertheless suggests that systematic offsets in masses could be partially arising due to the reddening corrections which have a stronger effect in UV. In the case of Balmer line studies, the mass offset is reduced by 40\,per cent when not accounting for the reddening in our fits. However, the systematic offsets of about 4\,per cent in \Teff\ is still being present in case of G11.\\

\item \textit{Flux calibration}: A systematically lower COS \Teff\ can result from calibration issues in the bluer end of the spectrum (<1200\,\AA). To verify this, we refitted the parameters cutting the blue edge of the spectrum i.e. considering the spectrum with $\lambda\geq1225$\,\AA\ which includes the red wing of the Ly$\alpha$. We find that the differences between the derived parameters are not significant, hence any COS calibration issue would need to affect all wavelengths equally.
    
Using \textit{HST} STIS and \textit{HST} WFC3, several recent studies have found good agreement between near-UV and optical Balmer line parameters \citep{Bohlin2014,Bohlin2019,Narayan2019,nicola2020,Axelrod2023}. However, these \textit{HST} spectrophotometric scales are calibrated using optical white dwarf parameters \citep{Bohlin2014}. More recently, \cite{2023cos..rept....5M} have done the re-calibration of \textit{HST} COS data by updating the CALSPEC standard models with \cite{Bohlin2020}, confirming that the re-calibrated data matche the models within 2\,per cent. As the re-calibrated COS data has been used in this work, this suggests that either 1) the re-calibration accuracy is closer to $\approx$4\,per cent (the offset seen in our study), or 2) that the white dwarf models in the far-UV \textit{HST} COS wavelength region have microphysics issues that are not present in the near-UV region observed from \textit{HST} STIS and \textit{HST} WFC3.
\end{enumerate}

\section{Conclusion}\label{sec:conc}
We conducted a large systematic study of \totalnumber\ DA white dwarfs for the first time by analysing the UV medium-resolution spectra obtained from \textit{HST} COS observations. The \Teff\ and \logg\ were derived by fitting the absolute fluxes of the sources with the updated white dwarf models by implementing two M-R relations (Montreal and La Plata Models), \textit{Gaia} EDR3 parallaxes and extinction values from STILISM/EXPLORE. The results from the two models suggest that different assumptions of H envelope compositions in M-R relations lead to differences in the measured masses of white dwarfs. The masses estimated from La Plata models are comparatively higher than Montreal models for stars with masses less than 0.6\,\Msun.

We carried out a comparative analysis of COS FUV parameters with previous spectroscopic and photometric studies to check the inconsistencies that can arise due to several reasons such as different models, fitting methods, and observed data. The effective temperatures from UV fits are found to be more consistent with the optical photometric studies (\textit{Gaia} and Pan-STARRS) with only a $\approx$1.5\,per cent systematic difference with COS values being lower. In comparison, COS UV fits are on average cooler by 3\,per cent compared to Balmer line fits. From the mass distribution study, we found that COS masses are systematically lower by 0.05\,\Msun\ than Balmer line values, while it reduces to 0.02\,\Msun\ in the case of the optical photometric studies. The smaller difference with photometric studies is expected given the similar fitting methods using the same \textit{Gaia} parallax values and similar atmospheric models. 

We argue that the systematic offsets are likely due to several reasons including i) uncertainties on the H envelope mass in the M-R relations, ii) issues in the Stark and neutral broadening theories affecting the Balmer and Lyman lines, iii) the effects of interstellar reddening which is stronger in UV than optical and iv) \textit{HST} COS flux calibration that is based on Balmer line white dwarf parameters. However, we have not reached a definitive conclusion over which is the dominant effect. Further investigations and efforts are necessary to understand the sources of these differences. We also have a spectroscopic sample of DB white dwarfs which have helium-dominated atmospheres observed under the \textit{HST} COS snapshot program. It will be interesting to check if a similar systematic offset is present between the COS UV and optical parameters of DBs like DAs, which we plan for a new study in the near future. 

Taking advantage of the comparisons of COS UV physical parameters with the optical studies, we identified 30 unresolved binary candidates. These candidates will be useful for constraining the white dwarf binary evolution models.  Hence, a detailed investigation and follow-up studies are required to confirm their binarity. We also find twelve objects with high RUWE where six of them show metal absorption lines in the COS spectra. The precise parameters obtained in this study will be useful for inferring their accurate metal abundances in order to understand metal pollution in white dwarfs.

\section*{Acknowledgements}
We thank Martin Barstow for providing the $FUSE$ spectra. This research is based on observations made with the NASA/ESA Hubble Space Telescope obtained from the Space Telescope Science Institute, which is operated by the Association of Universities for Research in Astronomy, Inc., under NASA contract NAS 5–26555. These observations are associated with programs 12169, 12474, 13652, 14077, 15073, 16011, and 16642. This research received funding from the European Research Council under the European Union’s Horizon 2020 research and innovation programme number 101002408 (MOS100PC) and 101020057 (WDPLANETS) the UK STFC consolidated grant ST/T000406/1. OT was supported by a FONDECYT project 321038.

This research made use of Astropy,\footnote{http://www.astropy.org} a community-developed core Python package for Astronomy \citep{astropy:2013, astropy:2018}, scipy \citep{2020SciPy-NMeth}, specutils \citep{nicholas_earl_2023_7803739}.

\section{Data availability}
The COS spectroscopy data underlying this paper are available in the raw form via the \textit{HST} MAST archive under the programs mentioned in the acknowledgements. 

\bibliographystyle{mnras}
\bibliography{ref}

\newpage
\appendix

\section{Observation details}
Table\,\ref{tab:obs_details} provides the details of the \textit{HST} COS snapshot survey spanning 2010$-$2023 with observation dates, exposure time, and the number of targets observed under each snapshot program.
\begin{table}
\centering
\caption{Observation details of \textit{HST}/COS snapshot survey of 307 DA white dwarfs.}
\addtolength{\tabcolsep}{-4pt}
\begin{tabular}{cl@{~--~}lcc}
\hline
Program ID & \multicolumn{2}{c}{Observation Date}  & Exp. time\,(s) & Observed \\
\hline
12169	&	2010 Sep 17 & 2011 Aug 30	&	400$-$1470 	&	54	\\
12474	&	2011 Oct 04 & 2013 Jul 02	&	600$-$1600	&	45	\\
13652	&	2014 Dec 01 & 2015 Jul 19	&	800$-$1600	&	30	\\
14077	&	2015 Oct 06 & 2017 Sep 28	&	800$-$1800	&	36	\\
15073	&	2017 Nov 04 & 2019 Oct 05	&	800$-$2000	&	78	\\
16011	&	2019 Nov 01 & 2020 Oct 03	&	1000$-$2000	&	19	\\
16642	&	2021 Dec 01 & 2023 Aug 02   &	1000$-$1800	&	109\\
\hline
\label{tab:obs_details}
\end{tabular}
\end{table}

\section{Atmospheric parameters of White Dwarfs}
\begin{figure*}
\centering
\includegraphics[width=\textwidth]{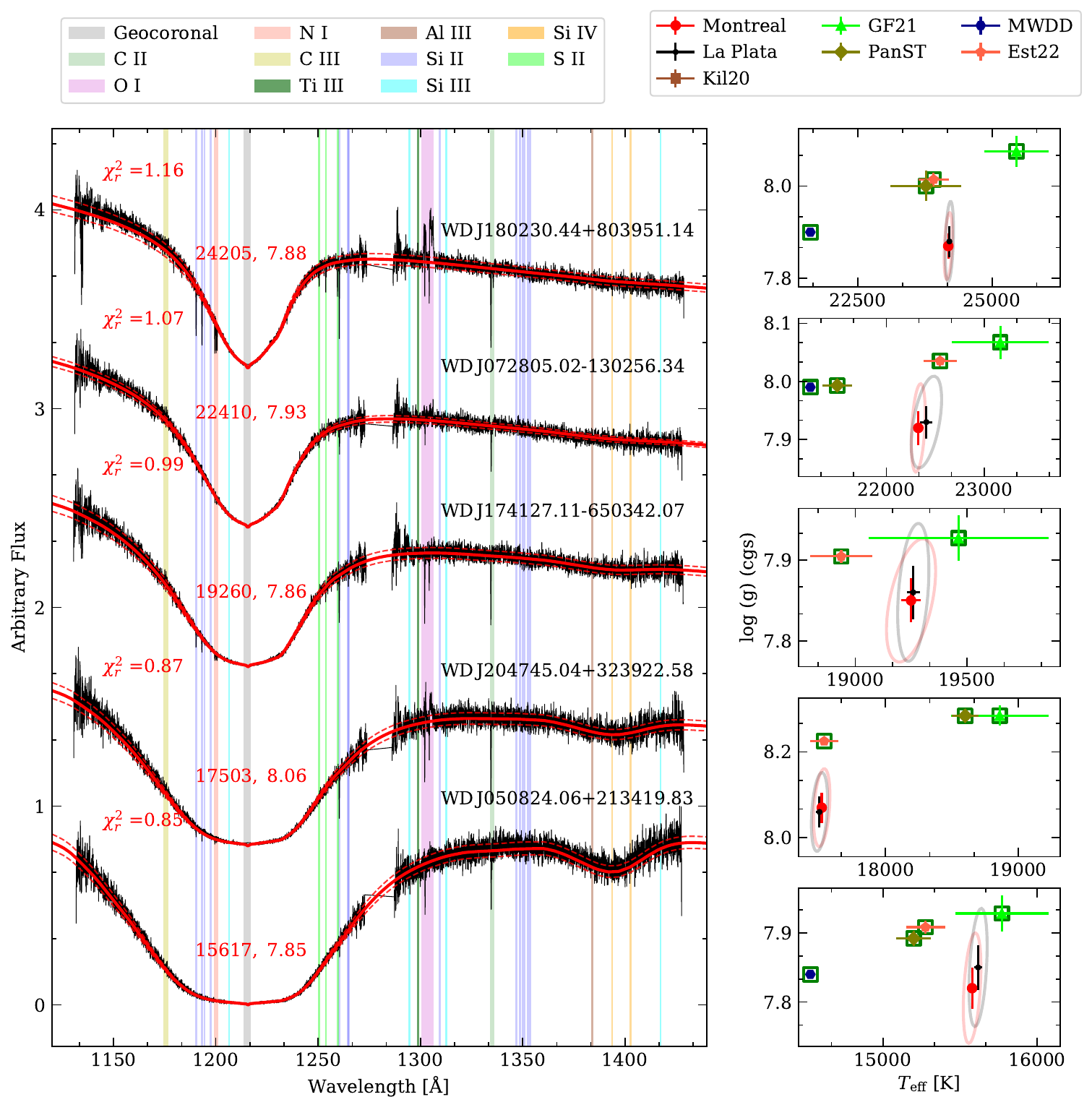}
\caption{Same as Fig.\,\ref{fig:spec_fit_wd_good} but for the white dwarfs where our COS data represent the first spectroscopic observations. The right panels show the photometric measurements from \textit{Gaia} \citep{nicola2021}, Est22 \citet{Esteban2022}, MWDD \citep{mwdd}, and Pan-STARRS in this work.}
\label{fig:spec_fit_wd_new}
\end{figure*}

Table\,\ref{tab:phot_params_wd} provides the COS atmospheric parameters of 49 white dwarf candidates discovered from \textit{Gaia}\,EDR3 \citep{nicola2021}. Four of these white dwarfs are shown in Fig.\,\ref{fig:spec_fit_wd_new}, with the model fits to the COS spectra in the left panels, and the atmospheric parameters measured from \textit{Gaia}, Pan-STARRS compared to those derived from the  COS data in the right panels. The full catalogue of \totalnumber\ white dwarfs is available online through Vizier. 

Table\,\ref{tab:params_wd_withism_lyman} presents the atmospheric parameters of 10 objects obtained with and without including the contribution of ISM Ly$\alpha$ in the models. Refer Sec.\,\ref{sec:outliers} for more details.

\begin{table*}
\centering
\caption{COS atmospheric parameters of 49 white dwarfs discovered by \textit{Gaia} (GF21) where the first spectroscopic observations were obtained as part of our \textit{HST} study. The subscript "Mont" and "LP" in \Teff, \logg, $M$ (mass), and $t$ (cooling age) denote the fit values obtained from Montreal and La Plata M-R relations respectively. The full table comprising details of \totalnumber\ white dwarfs is available online through Vizier.}
\addtolength{\tabcolsep}{-1pt}
\begin{tabular}{ccccccccccc}
\hline
Object & $T\rm{_{eff,Mont}}$ & $T\rm{_{eff,LP}}$ & $\log g\rm{_{Mont}}$ & $\log g\rm{_{LP}}$  & $M\rm{_{Mont}}$ & $M\rm{_{LP}}$ & Parallax & $t\rm{_{Mont}}$ & $t\rm{_{LP}}$ & $\chi^{2}_\mathrm{r}$\\
& \multicolumn{2}{c}{(K)} & \multicolumn{2}{c}{(dex)}  & \multicolumn{2}{c}{(M$_\odot$)} &(mas)& \multicolumn{2}{c}{(Myr)}\\
\hline
WD\	J002313.53+475259.55	&	18975	(38)	&	18963	(37)	&	7.96	(0.02)	&	7.95	(0.03)	&	0.60	(0.01)	&	0.59	(0.01)	&	13.83	(0.09/0.08)	&	76	&	78	&	0.96	\\
WD\	J003043.68+733738.23	&	19361	(34)	&	19475	(63)	&	7.78	(0.02)	&	7.82	(0.03)	&	0.51	(0.01)	&	0.54	(0.01)	&	11.32	(0.08/0.05)	&	46	&	59	&	1.04	\\
WD\	J004331.10+470134.30	&	20805	(29)	&	20772	(29)	&	8.26	(0.05)	&	8.25	(0.04)	&	0.78	(0.03)	&	0.76	(0.03)	&	18.68	(0.09/0.1)	&	110	&	109	&	1.06	\\
WD\	J022339.21+510454.25	&	17269	(46)	&	17399	(50)	&	7.69	(0.02)	&	7.75	(0.03)	&	0.46	(0.01)	&	0.50	(0.01)	&	13.41	(0.11/0.08)	&	65	&	83	&	0.87	\\
WD\	J023349.11$-$071534.01	&	22058	(42)	&	22163	(44)	&	7.75	(0.02)	&	7.79	(0.03)	&	0.50	(0.01)	&	0.53	(0.01)	&	13.14	(0.05/0.08)	&	25	&	33	&	1.14	\\
WD\	J030146.30+493659.64	&	15761	(17)	&	15805	(18)	&	7.78	(0.03)	&	7.81	(0.04)	&	0.50	(0.02)	&	0.52	(0.02)	&	17.40	(0.09/0.07)	&	113	&	130	&	0.85	\\
WD\	J045514.63$-$544145.41	&	17136	(23)	&	17195	(44)	&	7.78	(0.04)	&	7.80	(0.04)	&	0.50	(0.02)	&	0.52	(0.02)	&	19.92	(0.12/0.08)	&	80	&	93	&	0.95	\\
WD\	J050824.06+213419.83	&	15578	(23)	&	15617	(24)	&	7.82	(0.03)	&	7.85	(0.03)	&	0.52	(0.02)	&	0.54	(0.01)	&	19.42	(0.11/0.09)	&	128	&	145	&	0.85	\\
WD\	J055635.50$-$561006.57	&	21830	(69)	&	21781	(67)	&	8.15	(0.03)	&	8.14	(0.03)	&	0.72	(0.02)	&	0.70	(0.02)	&	13.11	(0.1/0.1)	&	67	&	68	&	1.08	\\
WD\	J063541.34$-$052430.64	&	21123	(21)	&	21165	(21)	&	7.84	(0.03)	&	7.86	(0.04)	&	0.54	(0.02)	&	0.56	(0.02)	&	18.17	(0.09/0.06)	&	34	&	43	&	1.34	\\
WD\	J072805.02$-$130256.34	&	22327	(29)	&	22410	(63)	&	7.92	(0.03)	&	7.93	(0.03)	&	0.59	(0.02)	&	0.59	(0.01)	&	13.65	(0.09/0.07)	&	31	&	35	&	1.07	\\
WD\	J073548.24+022423.49	&	22062	(27)	&	22101	(28)	&	7.82	(0.03)	&	7.83	(0.03)	&	0.53	(0.01)	&	0.55	(0.01)	&	12.40	(0.07/0.05)	&	27	&	35	&	1.06	\\
WD\	J081425.47$-$643211.05	&	18593	(52)	&	18592	(52)	&	7.94	(0.03)	&	7.94	(0.03)	&	0.59	(0.02)	&	0.59	(0.01)	&	16.77	(0.11/0.08)	&	80	&	84	&	0.97	\\
WD\	J082130.53$-$251140.78	&	20608	(33)	&	20695	(34)	&	7.77	(0.03)	&	7.80	(0.03)	&	0.51	(0.01)	&	0.53	(0.01)	&	10.35	(0.06/0.05)	&	34	&	45	&	0.97	\\
WD\	J082532.35$-$072823.21	&	15324	(14)	&	15349	(38)	&	7.90	(0.05)	&	7.91	(0.05)	&	0.56	(0.03)	&	0.57	(0.02)	&	28.19	(0.13/0.12)	&	158	&	168	&	0.90	\\
WD\	J083920.71$-$280132.44	&	25049	(21)	&	25056	(21)	&	7.89	(0.03)	&	7.89	(0.03)	&	0.58	(0.01)	&	0.58	(0.01)	&	10.42	(0.04/0.04)	&	17	&	21	&	1.13	\\
WD\	J091918.15$-$473354.38	&	23638	(53)	&	23615	(53)	&	7.93	(0.03)	&	7.92	(0.03)	&	0.59	(0.02)	&	0.59	(0.01)	&	12.31	(0.1/0.08)	&	23	&	27	&	1.15	\\
WD\	J094755.68$-$231234.10	&	22426	(37)	&	22463	(37)	&	7.83	(0.03)	&	7.84	(0.03)	&	0.54	(0.01)	&	0.55	(0.01)	&	10.45	(0.07/0.06)	&	26	&	32	&	0.99	\\
WD\	J104017.14$-$655324.81	&	21241	(29)	&	21299	(29)	&	7.83	(0.02)	&	7.85	(0.03)	&	0.54	(0.01)	&	0.56	(0.01)	&	10.19	(0.06/0.05)	&	33	&	41	&	1.09	\\
WD\	J105925.27$-$724409.93	&	19278	(31)	&	19282	(31)	&	7.90	(0.03)	&	7.90	(0.03)	&	0.57	(0.01)	&	0.57	(0.01)	&	13.06	(0.09/0.07)	&	61	&	67	&	0.97	\\
WD\	J121238.09$-$364240.22	&	19017	(30)	&	19250	(58)	&	7.62	(0.02)	&	7.70	(0.02)	&	0.44	(0.01)	&	0.49	(0.01)	&	12.04	(0.08/0.06)	&	39	&	65	&	1.00	\\
WD\	J144107.40$-$560154.83	&	21880	(79)	&	21946	(81)	&	7.85	(0.02)	&	7.87	(0.02)	&	0.55	(0.01)	&	0.56	(0.01)	&	10.24	(0.05/0.05)	&	30	&	37	&	1.03	\\
WD\	J150742.03$-$592754.43	&	22133	(27)	&	22135	(27)	&	7.94	(0.03)	&	7.94	(0.03)	&	0.60	(0.01)	&	0.59	(0.01)	&	12.37	(0.07/0.06)	&	33	&	38	&	1.08	\\
WD\	J165112.59$-$204106.36	&	20101	(45)	&	20037	(21)	&	7.98	(0.03)	&	7.96	(0.04)	&	0.61	(0.01)	&	0.60	(0.02)	&	16.64	(0.09/0.07)	&	61	&	62	&	1.01	\\
WD\	J170634.56$-$184047.13	&	20721	(68)	&	20808	(70)	&	7.87	(0.02)	&	7.89	(0.03)	&	0.56	(0.01)	&	0.57	(0.01)	&	11.69	(0.07/0.05)	&	40	&	48	&	1.02	\\
WD\	J174127.11$-$650342.07	&	19250	(45)	&	19260	(29)	&	7.85	(0.03)	&	7.86	(0.03)	&	0.54	(0.01)	&	0.55	(0.01)	&	14.85	(0.09/0.07)	&	55	&	64	&	0.99	\\
WD\	J174902.45$-$343255.27	&	19139	(49)	&	19057	(47)	&	8.41	(0.04)	&	8.38	(0.04)	&	0.87	(0.03)	&	0.84	(0.03)	&	18.10	(0.13/0.13)	&	198	&	194	&	0.90	\\
WD\	J175151.11$-$202308.72	&	18487	(48)	&	18440	(47)	&	8.12	(0.03)	&	8.11	(0.03)	&	0.69	(0.02)	&	0.68	(0.02)	&	18.47	(0.07/0.08)	&	124	&	126	&	1.03	\\
WD\	J175352.16+330622.62	&	16750	(31)	&	16783	(31)	&	7.88	(0.04)	&	7.90	(0.04)	&	0.55	(0.02)	&	0.57	(0.02)	&	28.01	(0.14/0.11)	&	108	&	120	&	1.03	\\
WD\	J175712.24+283957.46	&	18421	(40)	&	18443	(40)	&	7.89	(0.03)	&	7.90	(0.03)	&	0.56	(0.02)	&	0.57	(0.01)	&	15.34	(0.11/0.09)	&	74	&	82	&	0.89	\\
WD\	J180230.44+803951.14	&	24184	(31)	&	24205	(31)	&	7.87	(0.03)	&	7.88	(0.03)	&	0.57	(0.01)	&	0.57	(0.01)	&	11.87	(0.08/0.06)	&	19	&	24	&	1.16	\\
WD\	J180240.42$-$243603.86	&	18911	(34)	&	18935	(35)	&	7.87	(0.02)	&	7.87	(0.03)	&	0.55	(0.01)	&	0.56	(0.01)	&	14.37	(0.1/0.07)	&	62	&	70	&	1.02	\\
WD\	J180354.33$-$375202.95	&	18000	(26)	&	18000	(22)	&	7.90	(0.03)	&	7.90	(0.03)	&	0.56	(0.01)	&	0.57	(0.01)	&	16.58	(0.09/0.07)	&	84	&	91	&	1.01	\\
WD\	J181058.67+311940.94	&	15708	(29)	&	16077	(26)	&	7.11	(0.03)	&	7.28	(0.03)	&	0.27	(0.01)	&	0.35	(0.01)	&	20.44	(0.09/0.12)	&	13	&	121	&	1.27	\\
WD\	J182315.21+170639.42	&	20089	(25)	&	20171	(26)	&	7.76	(0.03)	&	7.79	(0.03)	&	0.50	(0.01)	&	0.53	(0.01)	&	13.13	(0.07/0.06)	&	38	&	49	&	1.05	\\
WD\	J184157.88+533818.93	&	20752	(60)	&	20957	(65)	&	7.63	(0.02)	&	7.70	(0.03)	&	0.45	(0.01)	&	0.50	(0.01)	&	11.17	(0.11/0.07)	&	28	&	44	&	1.03	\\
WD\	J184915.07$-$212603.48	&	21458	(54)	&	21542	(37)	&	7.75	(0.03)	&	7.78	(0.03)	&	0.50	(0.01)	&	0.53	(0.01)	&	10.05	(0.07/0.07)	&	28	&	37	&	1.05	\\
WD\	J191429.35$-$544019.71	&	25136	(18)	&	25124	(14)	&	8.22	(0.03)	&	8.21	(0.04)	&	0.76	(0.02)	&	0.75	(0.02)	&	17.10	(0.06/0.04)	&	40	&	41	&	1.30	\\
WD\	J191558.47$-$303535.44	&	17064	(21)	&	17081	(21)	&	7.87	(0.03)	&	7.88	(0.04)	&	0.55	(0.02)	&	0.56	(0.02)	&	19.37	(0.1/0.08)	&	98	&	108	&	0.91	\\
WD\	J191720.56+445239.38	&	21851	(70)	&	21817	(62)	&	8.25	(0.03)	&	8.24	(0.04)	&	0.78	(0.02)	&	0.76	(0.02)	&	12.99	(0.12/0.09)	&	88	&	87	&	0.98	\\
WD\	J192034.41$-$471529.44	&	18844	(50)	&	18847	(50)	&	7.93	(0.03)	&	7.93	(0.03)	&	0.58	(0.01)	&	0.58	(0.01)	&	14.47	(0.09/0.08)	&	73	&	78	&	0.94	\\
WD\	J192726.24+100710.03	&	24263	(68)	&	24319	(30)	&	7.75	(0.02)	&	7.77	(0.03)	&	0.51	(0.01)	&	0.53	(0.01)	&	12.62	(0.07/0.08)	&	17	&	23	&	1.31	\\
WD\	J193124.43+570419.66	&	22462	(51)	&	22432	(50)	&	8.01	(0.03)	&	8.00	(0.03)	&	0.64	(0.02)	&	0.62	(0.02)	&	14.40	(0.08/0.07)	&	37	&	40	&	1.17	\\
WD\	J193955.06+093219.39	&	21403	(69)	&	21398	(68)	&	7.92	(0.02)	&	7.92	(0.03)	&	0.58	(0.01)	&	0.58	(0.01)	&	12.15	(0.08/0.06)	&	38	&	43	&	0.94	\\
WD\	J204745.04+323922.58	&	17520	(25)	&	17503	(25)	&	8.07	(0.04)	&	8.06	(0.04)	&	0.66	(0.02)	&	0.65	(0.02)	&	16.74	(0.09/0.09)	&	136	&	138	&	0.87	\\
WD\	J210952.38+650721.93	&	20416	(28)	&	20403	(43)	&	8.26	(0.04)	&	8.25	(0.04)	&	0.78	(0.03)	&	0.76	(0.02)	&	16.07	(0.1/0.09)	&	118	&	117	&	0.97	\\
WD\	J214125.64$-$484953.75	&	15065	(33)	&	15076	(33)	&	7.94	(0.04)	&	7.94	(0.05)	&	0.58	(0.02)	&	0.58	(0.02)	&	23.58	(0.18/0.12)	&	180	&	187	&	0.88	\\
WD\	J220238.75$-$280942.13	&	20657	(31)	&	20625	(30)	&	8.26	(0.04)	&	8.25	(0.04)	&	0.78	(0.03)	&	0.76	(0.02)	&	15.38	(0.08/0.09)	&	113	&	112	&	0.94	\\
WD\	J230840.77$-$214459.60	&	15847	(37)	&	15881	(38)	&	7.92	(0.04)	&	7.93	(0.04)	&	0.57	(0.02)	&	0.58	(0.02)	&	30.51	(0.14/0.1)	&	145	&	154	&	0.95	\\
\hline
\label{tab:phot_params_wd}
\end{tabular}
\end{table*}

\begin{table*}
\centering
\caption{COS atmospheric parameters of 10 white dwarfs obtained with and without accounting for ISM Ly$\alpha$ in the models. The subscript "Mont" and "LP" in \Teff, \logg, M (mass), denote the fit values obtained from Montreal and La Plata M-R relations respectively.}
\addtolength{\tabcolsep}{-1pt}
\begin{tabular}{cccccccc}
\hline
Object & $T\rm{_{eff,Mont}}$ & $T\rm{_{eff,LP}}$ & $\log g\rm{_{Mont}}$ & $\log g\rm{_{LP}}$  & $M\rm{_{Mont}}$ & $M\rm{_{LP}}$ & $\chi^{2}_\mathrm{r}$\\
& \multicolumn{2}{c}{(K)} & \multicolumn{2}{c}{(dex)}  & \multicolumn{2}{c}{(M$_\odot$)} \\
\hline
\multicolumn{8}{c}{with ISM Ly$\alpha$}\\\hline
APASSJ085913.51-312416.3	&	30000	(	61 	)	&	30000	(	46	)	&	7.79	(	0.03	)	&	7.80	(	0.03	)	&	0.54	&	0.56	&	1.15	\\
HE1247-1130	&	27326	(	41 	)	&	27303	(	41	)	&	8.00	(	0.03	)	&	7.99	(	0.02	)	&	0.64	&	0.63	&	1.16	\\
HE2345-4810	&	28773	(	47 	)	&	29091	(	45	)	&	7.31	(	0.02	)	&	7.40	(	0.02	)	&	0.37	&	0.43	&	1.18	\\
PG1513+442	&	28507	(	44 	)	&	28493	(	44	)	&	7.95	(	0.03	)	&	7.95	(	0.03	)	&	0.62	&	0.61	&	1.16	\\
PG1620+260	&	27642	(	83 	)	&	27787	(	87	)	&	7.67	(	0.02	)	&	7.71	(	0.02	)	&	0.48	&	0.52	&	1.15	\\
WD1412-109	&	25287	(	26 	)	&	25266	(	26	)	&	7.97	(	0.02	)	&	7.96	(	0.02	)	&	0.62	&	0.61	&	1.17	\\
WD1451+006	&	25621	(	31 	)	&	25661	(	32	)	&	7.88	(	0.03	)	&	7.88	(	0.03	)	&	0.57	&	0.58	&	1.12	\\
WDJ152310.59+305344.80	&	25045	(	21 	)	&	25093	(	21	)	&	7.77	(	0.02	)	&	7.79	(	0.03	)	&	0.52	&	0.54	&	1.14	\\
WDJ155501.99+351328.65	&	26499	(	35 	)	&	26527	(	36	)	&	7.84	(	0.03	)	&	7.85	(	0.03	)	&	0.56	&	0.57	&	1.35	\\
WDJ170909.53+473134.68	&	24790	(	98 	)	&	24776	(	98	)	&	7.94	(	0.02	)	&	7.93	(	0.03	)	&	0.60	&	0.59	&	1.21	\\\hline
\multicolumn{8}{c}{without ISM Ly$\alpha$}\\\hline
APASSJ085913.51-312416.3	&	28930	(	68 	)	&	29000	(	55	)	&	7.70	(	0.03	)	&	7.73	(	0.03	)	&	0.50	&	0.53	&	1.46	\\
HE1247-1130	&	27063	(	36 	)	&	27044	(	36	)	&	7.97	(	0.03	)	&	7.96	(	0.03	)	&	0.62	&	0.61	&	1.23	\\
HE2345-4810	&	28317	(	49 	)	&	28646	(	54	)	&	7.26	(	0.01	)	&	7.37	(	0.02	)	&	0.35	&	0.43	&	1.25	\\
PG1513+442	&	28000	(	34 	)	&	28000	(	34	)	&	7.90	(	0.03	)	&	7.90	(	0.03	)	&	0.59	&	0.59	&	1.28	\\
PG1620+260	&	27001	(	30 	)	&	27131	(	32	)	&	7.61	(	0.02	)	&	7.66	(	0.02	)	&	0.46	&	0.50	&	1.26	\\
WD1412-109	&	24515	(	91 	)	&	24551	(	92	)	&	7.87	(	0.02	)	&	7.88	(	0.02	)	&	0.57	&	0.57	&	1.38	\\
WD1451+006	&	25572	(	31 	)	&	25605	(	32	)	&	7.87	(	0.03	)	&	7.88	(	0.03	)	&	0.57	&	0.58	&	1.21	\\
WDJ152310.59+305344.80	&	24567	(	88 	)	&	25031	(	21	)	&	7.71	(	0.02	)	&	7.78	(	0.03	)	&	0.49	&	0.54	&	1.20	\\
WDJ155501.99+351328.65	&	26204	(	30 	)	&	26259	(	31	)	&	7.80	(	0.03	)	&	7.82	(	0.03	)	&	0.54	&	0.55	&	1.42	\\
WDJ170909.53+473134.68	&	24416	(	36 	)	&	24422	(	36	)	&	7.89	(	0.03	)	&	7.89	(	0.03	)	&	0.58	&	0.58	&	1.36	\\
 \hline
\label{tab:params_wd_withism_lyman}
\end{tabular}
\end{table*}

\section{Comparison of COS parameters with other literature}
Tables\,\ref{tab:K09u_params}, \ref{tab:iue_params}, and \ref{tab:panst_params} present the atmospheric parameters obtained from fitting the  SPY spectra \citep{Koester2009} with updated TB09 profiles, $IUE$ spectra, and, Pan-STARRS photometry, respectively.

\begin{table*}
\centering
\caption{Atmospheric parameters of 123 objects \citep{Koester2009} obtained from the analysis of SPY spectra using updated TB09 profiles. The first 10 rows are shown for illustration, the full table is available online through Vizier.}
\begin{tabular}{ccccc} \hline
Object & \Teff & \logg & S/N & $\rm{\chi^2}$ \\\hline
HE\	0131+0149	&	14792	(55)	&	7.87	(0.01)	&	21.7	&	1.01	\\
HE\	0305-1145	&	26939	(103)	&	7.83	(0.02)	&	17.4	&	3.21	\\
HE\	0308-2305	&	23989	(50)	&	8.63	(0.01)	&	30.6	&	1.82	\\
HE\	0358-5127	&	23389	(83)	&	8.03	(0.01)	&	20.1	&	1.37	\\
HE\	0403-4129	&	22466	(103)	&	7.99	(0.02)	&	14.7	&	1.43	\\
HE\	0414-4039	&	21089	(133)	&	8.16	(0.02)	&	12.4	&	1.84	\\
HE\	0416-1034	&	24854	(56)	&	7.99	(0.01)	&	32.6	&	1.27	\\
HE\	0418-1021	&	22893	(39)	&	8.45	(0.01)	&	34.0	&	1.74	\\
HE\	0418-5326	&	27133	(90)	&	7.92	(0.02)	&	17.5	&	1.17	\\
HE\	0452-3444	&	20647	(59)	&	7.93	(0.01)	&	21.2	&	3.68	\\
\hline
\label{tab:K09u_params}
\end{tabular}
\end{table*}

\begin{table*}
\centering
\caption{Atmospheric parameters derived from \textit{IUE} observations for 15 stars in common with the COS survey, used for a comparative analysis. The parameters are obtained using La Plata M-R relation. The parameters are provided for two cases: one obtained using the full $IUE$ spectrum while the other considering only the spectral region corresponding to the COS wavelength range (1150--1430\,\AA).}
\begin{tabular}{ccccccc} \hline
& \multicolumn{3}{c}{Full spectrum} & \multicolumn{3}{c}{spectral range (1150--1430\,\AA)} \\
Object	& 	\Teff	&	\logg	&	$\chi_r^2$ 	&	\Teff	&	\logg & $\chi_r^2$ 	\\\hline
PG\ 1143+321	&	16139	(803)	&	8.11	(0.29)	&	1.65	&	15915	(161)	&	8.05	(0.08)	&	1.51	\\
WD\ 0047$-$524	&	18155	(378)	&	7.79	(0.15)	&	2.28	&	18361	(146)	&	7.83	(0.07)	&	2.01	\\
WD\ 0231$-$054	&	13117	(425)	&	8.50	(0.34)	&	1.36	&	12965	(106)	&	8.47	(0.08)	&	3.00	\\
WD\ 0232+525	&	16586	(848)	&	8.19	(0.32)	&	1.35	&	16981	(149)	&	8.28	(0.06)	&	1.63	\\
WD\ 0348+339	&	13823	(97)	&	8.35	(0.10)	&	1.5	&	14405	(160)	&	8.56	(0.10)	&	2.15	\\
WD\ 0406+169	&	15795	(1008)	&	8.45	(0.42)	&	2.05	&	15368	(182)	&	8.34	(0.11)	&	2.44	\\
WD\ 0410+117	&	20294	(205)	&	7.93	(0.05)	&	2.26	&	20442	(80)	&	7.95	(0.03)	&	1.96	\\
WD\ 1052+273	&	22692	(1624)	&	8.42	(0.38)	&	1.24	&	22340	(190)	&	8.37	(0.08)	&	1.39	\\
WD\ 1104+602	&	18098	(273)	&	8.09	(0.12)	&	2.55	&	18721	(135)	&	8.22	(0.06)	&	2.26	\\
WD\ 1327$-$083	&	14250	(960)	&	7.88	(0.53)	&	0.94	&	14569	(135)	&	7.99	(0.08)	&	1.12	\\
WD\ 1713+695	&	16563	(2006)	&	8.23	(0.80)	&	0.81	&	16030	(325)	&	8.09	(0.18)	&	1.17	\\
WD\ 1919+145	&	14321	(397)	&	7.91	(0.21)	&	2.18	&	15235	(85)	&	8.20	(0.04)	&	2.36	\\
WD\ 2047+372	&	13846	(171)	&	8.02	(0.11)	&	1.51	&	14750	(85)	&	8.34	(0.06)	&	2.01	\\
WD\ 2126+734	&	15577	(549)	&	7.92	(0.23)	&	1.57	&	16062	(126)	&	8.06	(0.07)	&	2.46	\\
WD\ 2341+322	&	12301	(77)	&	7.84	(0.09)	&	1.57	&	12660	(47)	&	8.03	(0.04)	&	1.62	\\

\hline
\label{tab:iue_params}
\end{tabular}
\end{table*}

\begin{table*}
\centering
\caption{Atmospheric parameters of 257 stars obtained using Pan-STARRS photometry. The first 10 rows are shown for illustration, the full catalogue is available online through Vizier.}
\begin{tabular}{cccc} \hline
Object & \Teff & \logg & Mass \\ \hline
APASS\ J013001.36+263857.4	&	14216	(265)	&	8.19	(0.02)	&	0.72	\\
APASS\ J081237.87+173700.3	&	15207	(79)	&	8.00	(0.01)	&	0.61	\\
APASS\ J083857.48-214611.0	&	21274	(435)	&	7.92	(0.03)	&	0.58	\\
APASS\ J085913.51-312416.3	&	11334	(93)	&	7.13	(0.02)	&	0.29	\\
APASS\ J090028.59-090923.2	&	19796	(164)	&	7.80	(0.01)	&	0.52	\\
APASS\ J145521.26+565544.3	&	14907	(142)	&	7.96	(0.01)	&	0.59	\\
APASS\ J151754.65+103043.7	&	19607	(310)	&	7.89	(0.03)	&	0.56	\\
APASS\ J152827.83-251503.0	&	15252	(102)	&	8.35	(0.01)	&	0.83	\\
APASS\ J195622.94+641358.0	&	14516	(128)	&	7.52	(0.01)	&	0.41	\\
APASS\ J202336.88-111551.3	&	15856	(108)	&	7.95	(0.01)	&	0.59	\\
\hline
\label{tab:panst_params}
\end{tabular}
\end{table*}

\begin{figure*}
\centering
\includegraphics[width=\textwidth]{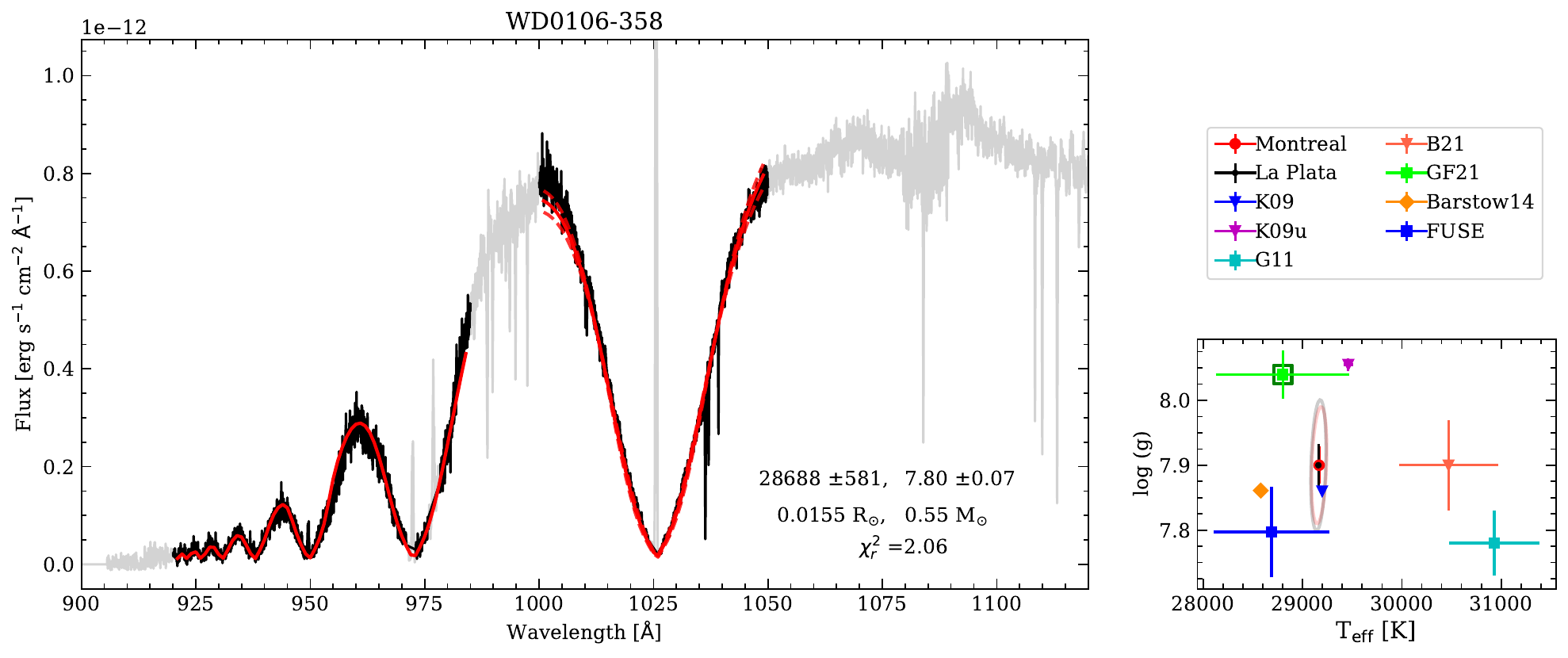}
\caption{WD model fit (red solid line) to the \textit{FUSE} spectrum of WD\,0106$-$358. Only the spectral regions covering Lyman lines (shown in black) were included in the fit. The best-fit parameters obtained from the fitting are labeled in the left panel and shown as a blue square in the right panel. The parameters from \citet{Barstow2014} are shown as orange diamond on the left panel. Refer Fig.\,\ref{fig:spec_fit_wd_good} for more details on the labels.}
\label{fig:fuse_fit}
\end{figure*}

\citet{Barstow2014} measured atmospheric parameters of 89 DA white dwarfs spanning \Teff\ range 20\,000$-$77\,000\,K from \textit{FUSE} observations covering wavelength region 912$-$1180\,\AA, which includes all the lines of the Lyman series. To check whether fitting the Lyman series gives consistent results with those of our COS and \textit{IUE} analyses, which covered only a single Lyman line, we fitted the calibrated \textit{FUSE} spectra of the three objects in common between our COS observations and the \textit{FUSE} sample of \citet{Barstow2014}. We adopted the same procedure as for fitting the COS spectra, masked the geo-coronal lines, and considered only the spectral regions covering the Lyman lines (1000$-$1050\,\AA\, for Ly$\beta$ and 920$-$985\,\AA\, for the higher Lyman lines) to avoid regions affected by instrumental artefacts or numerous photospheric metal lines. Fig.\,\ref{fig:fuse_fit} illustrates the fit to the \textit{FUSE} spectrum of WD\,0106$-$358. The uncertainties associated with the fit parameters are determined by averaging the values of \Teff\ and \logg, which are obtained by independently fitting the two spectral regions covering the Lyman lines. We found that the resulting \Teff\ and \logg\ are in good agreement (within 3$\sigma$) with the parameters reported by \citet{Barstow2014} and derived from the COS data in this work. It is worth noting that spectroscopic analyses, including those of \citet{Barstow2014}, carried out prior to the availability of \textit{Gaia} parallaxes, were subject to correlations between \Teff\ and \logg. Fitting the space-based, flux-calibrated COS, \textit{IUE}, and \textit{FUSE} spectra largely removes this correlation and leads to consistent results across the different instruments.

Figs.\,\ref{fig:spec_uv_comp_li05}, \ref{fig:phot_uv_Ki20}, and, \ref{fig:phot_uv_est22} show the comparisons of COS atmospheric parameters (\Teff\ and \logg) with \cite{Liebert2005} based on Balmer line fits, \cite{Kilic2020} based on SDSS ($u$) and pan-STARRS ($grizy$) photometry, and, {\cite{Esteban2022} based on \textit{Gaia} DR3 data respectively. See Table\,\ref{tab:compare_table} and Sect.\,\ref{sec:comp} for more details.

\begin{figure}
\centering
\includegraphics[width=\columnwidth]{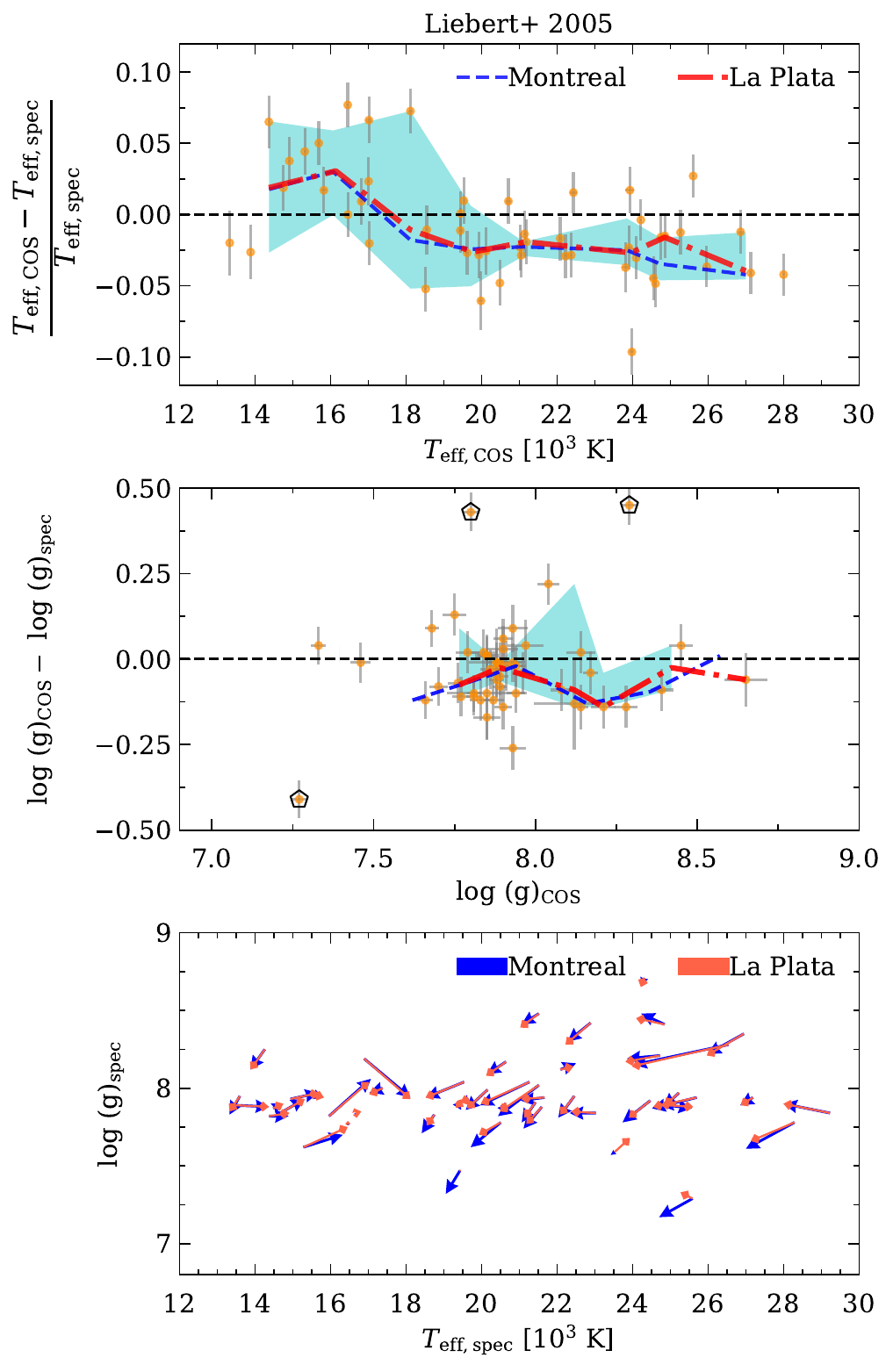}
\caption{\Teff\ and \logg\ differences of UV estimates with \citet{Liebert2005} in top and middle panels, respectively. The bottom panel shows the correlation between them. For a description of symbols, refer to Fig.\,\ref{fig:spec_uv_comp_k14}.}
\label{fig:spec_uv_comp_li05}
\end{figure}

\begin{figure}
\centering
\includegraphics[width=\columnwidth]{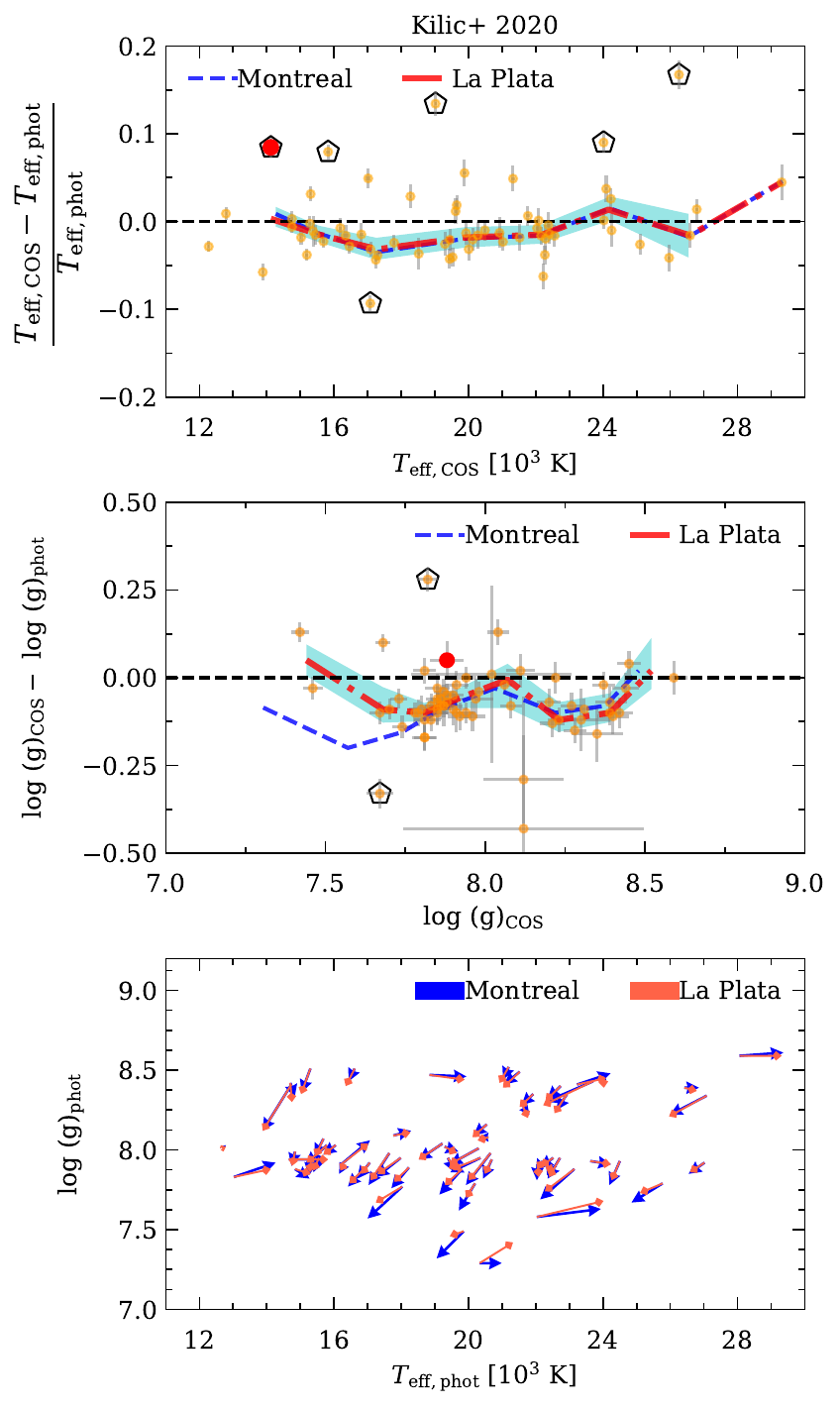}
\caption{\Teff\ and \logg\ differences of UV estimates with \citet{Kilic2020} in top and middle panels, respectively. The bottom panel shows the correlation between them. For a description of symbols, refer to Fig.\,\ref{fig:spec_uv_comp_k14}.}
\label{fig:phot_uv_Ki20}
\end{figure}

\begin{figure}
\centering
\includegraphics[width=\columnwidth]{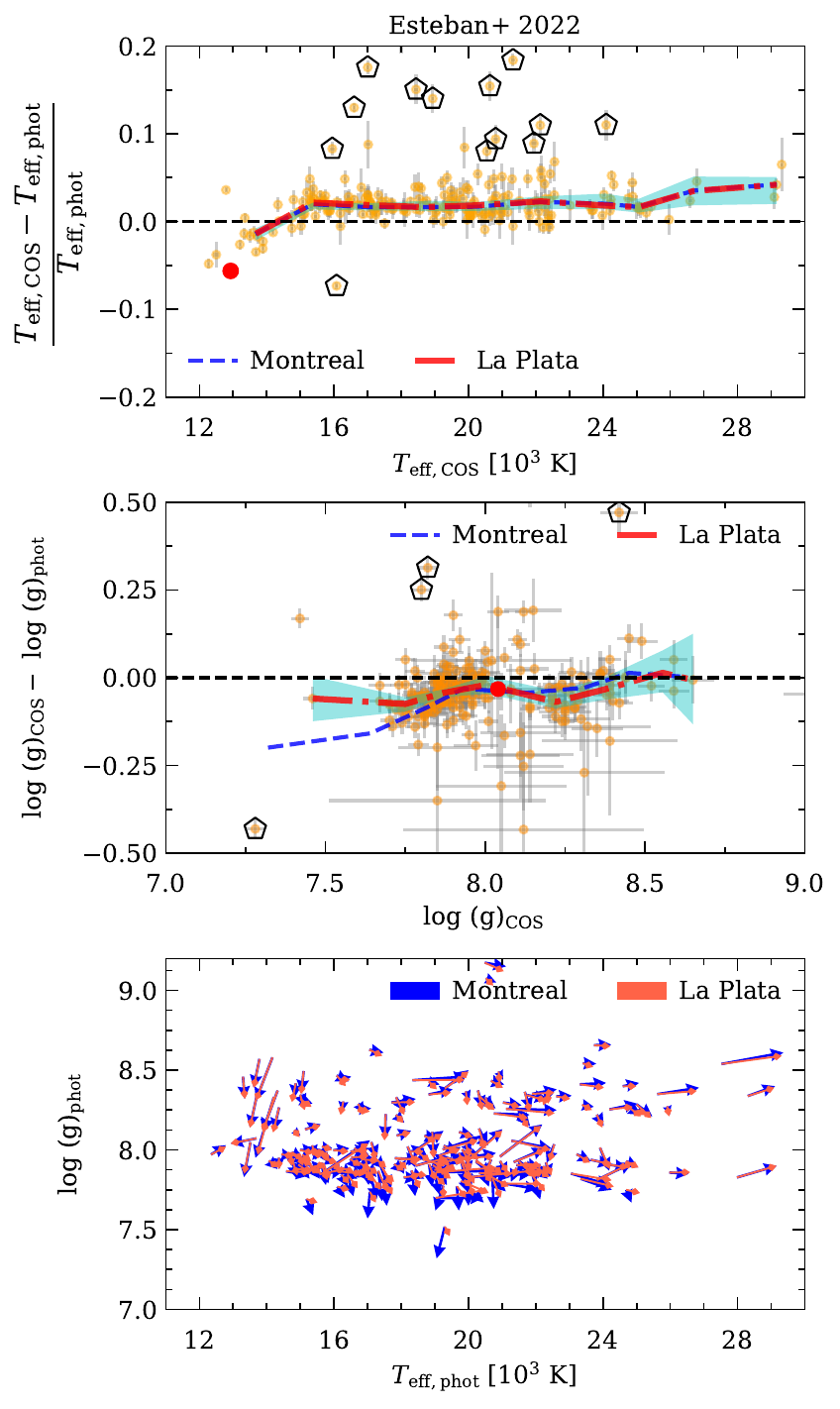}
\caption{\Teff\ and \logg\ differences of UV estimates with \citet{Esteban2022} in top and middle panels, respectively. The bottom panel shows the correlation between them. For a description of symbols, refer to Fig.\,\ref{fig:spec_uv_comp_k14}.}
\label{fig:phot_uv_est22}
\end{figure}

\end{document}